\documentclass[12pt,preprint]{emulateapj}
\usepackage{natbib}

\slugcomment{To appear in ApJ}

\shorttitle{SONYC: NGC1333}
\shortauthors{Scholz et al.}

\begin{document}
\bibliographystyle{apj}


\title{Substellar Objects in Nearby Young Clusters (SONYC) IV:\\ 
A census of very low mass objects in NGC1333}


\author{Alexander Scholz\altaffilmark{1}, Koraljka Muzic\altaffilmark{2}, 
Vincent Geers\altaffilmark{3}, Mariangela Bonavita\altaffilmark{2}, Ray Jayawardhana\altaffilmark{2,**}, 
Motohide Tamura\altaffilmark{4}}

\email{aleks@cp.dias.ie}

\altaffiltext{1}{School of Cosmic Physics, Dublin Institute for Advanced Studies, 31 Fitzwilliam Place,
Dublin 2, Ireland}
\altaffiltext{2}{Department of Astronomy \& Astrophysics, University of Toronto, 50 St. George Street, Toronto, 
ON M5S 3H4, Canada}
\altaffiltext{3}{Institute for Astronomy, ETH Zurich, Wolfgang-Pauli-Strasse 27, 8093 Zurich, Switzerland}
\altaffiltext{4}{National Astronomical Observatory, Osawa 2-21-2, Mitaka, Tokyo 181, Japan}
\altaffiltext{**}{Principal Investigator of SONYC}

\begin{abstract}
SONYC -- {\it Substellar Objects in Nearby Young Clusters} -- is a program to investigate the 
frequency and properties of young substellar objects with masses down to a few times that of Jupiter. 
Here we present a census of very low mass objects in the $\sim 1$\,Myr old cluster NGC1333. We 
analyze near-infrared spectra taken with FMOS/Subaru for 100 candidates from our deep, wide-field 
survey and find 10 new likely brown dwarfs with spectral types of M6 or later. Among them, there 
are three with $\gtrsim$M9 and one with early L spectral type, corresponding to masses of 0.006 to 
$\lesssim 0.02\,M_{\odot}$, so far the lowest mass objects identified in this cluster. The combination 
of survey depth, spatial coverage, and extensive spectroscopic follow-up makes NGC1333 one of the most
comprehensively surveyed clusters for substellar objects. In total, there are now 51 objects with 
spectral type M5 or later and/or effective temperature of 3200\,K or cooler identified in NGC1333; 
30-40 of them are likely to be substellar. NGC1333 harbours about half as many brown dwarfs as stars, 
which is significantly more than in other well-studied star forming regions, thus raising the possibility
of environmental differences in the formation of substellar objects. The brown dwarfs in NGC1333 are 
spatially strongly clustered within a radius of $\sim 1$\,pc, mirroring the distribution of the stars. 
The disk fraction in the substellar regime is $<66$\%, lower than for the total population (83\%) but 
comparable to the brown dwarf disk fraction in other 2-3\,Myr old regions.
\end{abstract}

\keywords{stars: circumstellar matter, formation, low-mass, brown dwarfs -- planetary systems}

\section{Introduction}
\label{s1}

Brown dwarfs are objects with masses too low to sustain stable hydrogen burning
($M<0.08\,M_{\odot}$) and as such intermediate in mass between low-mass stars and giant
planets \citep{2000prpl.conf.1313O}. The substellar mass regime is crucial to test how the 
physics of the formation and early evolution of stars depends on object mass, thus may 
help address some of the most relevant issues in this field. One example is the origin 
of the initial mass function (IMF) and the relative importance of dynamical interactions, 
fragmentation, and accretion in setting the mass of objects \citep{2007prpl.conf..149B}. 

Surveys in star forming regions indicate that the mass regime of free-floating 
brown dwarfs extends down to masses below the Deuterium burning limit at $0.015\,M_{\odot}$ 
\citep[e.g.][]{2000Sci...290..103Z,2000MNRAS.314..858L}, i.e. it is overlapping with 
the domain of massive planets. The currently available surveys, however, are not complete
in the substellar regime. Only small regions have been observed with sufficient depth
to detect the lowest-mass brown dwarfs. Moreover, in many cases the brown dwarf candidates
are selected based on their mid-infrared excess and the presence of disks 
\citep[e.g.][]{2006ApJ...644..364A}, which introduces an obvious bias. In other cases, the presence 
of methane absorption is used to identify objects \citep[e.g.][]{2009A&A...508..823B}, which 
only finds T dwarfs in a limited temperature regime.

\begin{figure}
\center
\includegraphics[width=8.5cm]{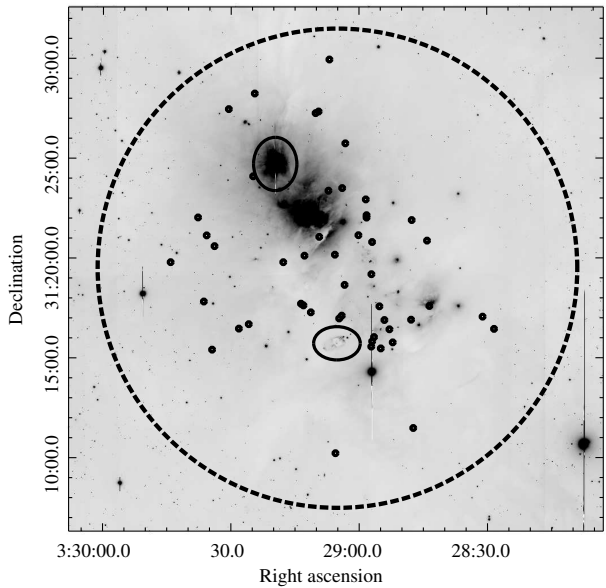} 
\caption{Subaru/Suprime-Cam i-band image for our target region NGC1333 in the Perseus star forming region. 
Marked are the cluster radius with a large dashed circle, the two well-known objects BD+30 549 
(north) and HH 7-11 (south) with small ellipses, and the population of very low mass members
(see this paper, Tables \ref{t1} and \ref{t2}) with small circles. The reflection nebula
NGC1333 is slightly north of the image center. Coordinates are J2000. For more information on 
this image, see \citet{2009ApJ...702..805S}. \label{f50}}
\end{figure}

In the SONYC project (short for: Substellar Objects in Nearby Young Clusters) we aim for 
a more complete census of brown dwarfs in star forming regions. For a number of regions, we 
have carried out deep photometric surveys in the optical and near-infrared, to facilitate 
a primary candidate selection based on photospheric colours. This is complemented by Spitzer
data to identify additional objects with disks. We have published the photometric survey as
well as follow-up spectroscopy for three regions so far: NGC1333 \citep{2009ApJ...702..805S}, 
$\rho$\,Oph \citep{2011ApJ...726...23G}, and Chamaeleon-I \citep{2011ApJ...732...86M}. A 
fourth paper with additional spectroscopy in $\rho$\,Oph is submitted (Muzic et al.).
Based on these results, the largest population of brown dwarfs is found in NGC1333,
a very young ($\sim 1$\,Myr old) cluster in the Perseus OB2 association at a distance
of $\sim 300$\,pc \citep{1996AJ....111.1964L,1999AJ....117..354D,2002A&A...387..117B}.
Fig. \ref{f50} shows a deep optical image of the cluster NGC1333 with some of the 
relevant features marked.

Here we present new follow-up spectroscopy in NGC1333 for a large sample of additional 
candidates from our photometric survey (Sect. \ref{s2}) and identify 10 new, previously unknown 
very low mass members (Sect. \ref{s3}). Combining these with the known members yields 51 objects 
with spectral type $\ge $M5 in this cluster. In Sect. \ref{s4} we analyse the brown dwarf census 
for NGC1333, including the mass function, the spatial distribution and the disk properties. The 
conclusions are given in Sect. \ref{s5}. Throughout this paper, we make use of a large number
of different samples for objects in the area of NGC1333. In Table \ref{t15} we provide
an overview of the most important samples and link them to the corresponding sections
of the paper.

\begin{deluxetable}{ll}
\tabletypesize{\scriptsize}
\tablecaption{Overview of the various samples in NGC1333 used in this paper
\label{t15}}
\tablewidth{0pt}
\tablehead{
\colhead{Sample} & \colhead{No.}}
\tablecolumns{2}
\startdata
Objects observed with FMOS  (Sect. \ref{s3})                  & 100\\
\hspace{0.5cm} excluded as very low mass sources               & 63\\
\hspace{0.5cm} confirmed as young very low mass sources (YVLM) & 26\\
\hspace{1.0cm} newly identified                                & 10\\
\hline
Objects with spectral type $>$M5 (Sect. \ref{s4})              & 51\\
\hspace{0.5cm} identified in this paper                        & 10\\
\hspace{0.5cm} identified in \citet{2009ApJ...702..805S}       & 20\\
\hspace{0.5cm} with estimated masses $<0.08\,M_{\odot}$        & 30-40\\
\hspace{0.5cm} with Spitzer counterpart (Sect. \ref{s44})      & 41\\
\hspace{1.0cm} with mid-infrared excess at 3-8$\,\mu m$        & 27\\
\hline
Candidates selected from the iz diagram (Sect. \ref{s41})      & 196\\
\hspace{0.5cm} with spectroscopy from MOIRCS or FMOS           & 98\\
\hspace{0.5cm} confirmed by our spectra                        & 27\\
\hspace{0.5cm} confirmed by other groups                       & 8\\
\hline
Class I and II sources in NGC1333\tablenotemark{a} (Sect. \ref{s42}) & 137\\
\hspace{0.5cm} with 2MASS detection                            & 94\\
\hspace{0.5cm} with estimated masses $0.02<M<0.08\,M_{\odot}$  & 29\\
\hspace{1.0cm} corrected for disk fraction                     & 35
\enddata
\tablenotetext{a}{\citet{2008ApJ...674..336G}}
\end{deluxetable}

\section{Observations and data reduction}
\label{s2}

Our spectra were obtained with the Fiber Multi Objects Spectrograph (FMOS) at the Subaru Telescope
\citep{kimura10} on the night of 2010 November 27. Four-hundred fibers, each of 1.2" diameter are 
configured by the fiber positioner system of FMOS in the 30' diameter field of view, with an
accuracy of 0.2" rms. The patrol radius of each spine is 87", while the minimum 
spacing between two neighboring spines is 12". 

The spectra are extracted by the two spectrographs (IRS1 and IRS2). Our data were obtained in 
shared-risk mode, using only one of the spectrographs (IRS1, 200 fibers). IRS1 is equipped 
with a $2048 \times 2048$ pixel HAWAII-II HgCdTe detector, and a mask mirror for OH airglow 
supression. With the low-resolution mode ($R\sim 600$) the spectrograph yields a coverage from 
0.9 to 1.8$\,\mu m$ (J- and H-bands). We obtained 10 exposures with 15 min on-source 
time each.

The observations were carried out in the Normal Beam Switching (NBS) mode, i.e. the same amount 
of time was spent to observe the sky, which is achieved by offseting the telescope by 10-15". 
The seeing during the science observations was stable at $\sim 1$". Since no stars suitable for 
telluric correction are found within our science field-of-view, we observed several standard 
star fields, covering the range of airmasses at which the science target was observed 
(airmass between 1-2). The observed standard stars have spectral types in the range F4 - G5.

Data reduction was carried out using the data reduction package supplied by the Subaru staff. 
The package consists of IRAF tasks and C programs using the CFITSIO library. The data reduction 
package contains the tasks for standard reduction of NIR spectra, performing sky subtraction, 
bad-pixel and flat-field correction, wavelength calibration, flux calibration and 
telluric correction. In this last step, each 15-minute exposure was calibrated using a standard 
star at the appropriate airmass. Finally, the ten individual exposures were averaged. 
For the analysis (Sect. \ref{s3}) the spectra were binned to a uniform wavelength interval of 
5\,\AA~and smoothed with a small-scale median filter. For the reduced spectra, the signal-to-noise 
ratio in the H-band ranges from 10 to 70.

In total, we covered 100 targets, from which 71 are selected from our IZ candidate catalogue 
\citep{2009ApJ...702..805S}. 10 additional targets have been selected by combining our JK-catalogue 
with the 'HREL' catalogue from the Spitzer `Cores to Disk' (C2D) Legacy program \citep{2009ApJS..181..321E}. 
All 10 have colour-excess in IRAC bands and thus should have a disk (see Sect. \ref{s45}). 19 fibres 
were placed on known M-type members for reference 
\citep[from][]{2004AJ....127.1131W,2007AJ....133.1321G,2009ApJ...702..805S,2009AJ....137.4777W}.

To test for possible effects of imperfect calibration, we compared the H-band spectra for six
objects observed with MOIRCS \citep{2009ApJ...702..805S} and with FMOS (this paper). For three of them 
there is excellent agreement (SONYC-NGC1333-1, 5, 8), while for the others there are slight differences 
in the spectral slope. Using the method outlined in Sect. \ref{s32}, we 
measured the spectral types for the four out of six MOIRCS spectra which cover the entire H-band. All 
four give types that are later than those derived from FMOS, by 0.4, 1.3, 0.5, 0.7 subtypes, i.e. the
differences are larger than our internal accuracy of 0.4 subtypes.

These discrepancies may indicate residual problems with the telluric calibration in the MOIRCS and/or 
the FMOS data. These problems could be induced by the use of multi-object facilities: Since we stay on the
target field for long integration times and the fields themselves do not cover adequate telluric standards,
there is a significant time and position offset between science targets and standards, i.e. the depth of
the telluric bands could potentially be quite different between science and standard fields. 

Stable conditions, as they were present for the FMOS observations, should mitigate this effect. The
FMOS data also have the wider wavelength coverage, which facilitates the telluric correction and the
spectral analysis. Therefore we put more trust in the quantities derived from FMOS spectra. 

\section{Spectral analysis} 
\label{s3}

\subsection{Selection of young very low mass objects}
\label{s31}

In total we obtained spectra for 100 objects with FMOS. Based on the broadband
spectral shape in the near-infrared, young very low mass sources can be reliably separated from 
more massive stars. Objects with very low masses and thus effective temperature below $\sim 3500$\,K or
spectral types of $\sim$M3 or later show broad absorption features of H$_2$O, which distinguishes them clearly
from the smooth near-infrared slopes of more massive and hotter stars \citep{2005ApJ...623.1115C}. The depth 
of these absorption troughs is a strong function of effective temperature.  

The most important spectral feature for our purposes is the H-band `peak', formed by the two H$_2$O absorption 
bands at 1.3-1.5 and 1.75-2.05$\,\mu m$. The shape of this feature is gravity sensitive; while it appears round 
with a flat top in evolved field objects, it is triangular with a pronounced peak at 1.67$\,\mu m$ in young objects
with ages of $\lesssim 100$\,Myr \citep{2006ApJ...639.1120K,2006ApJ...653L..61B,2010A&A...519A..93B}. In addition,
H$_2$O absorption causes a sharp edge at 1.35$\,\mu m$ and another `peak' in the K-band. 

We use these features to identify young very low mass sources in the FMOS sample. We are looking for 
objects showing structure over the full spectral regime, as opposed to a smooth slope. In particular, we select
objects with a) a pointy peak in the H-band and b) an absorption edge at 1.35$\,\mu m$. Out of 100 
FMOS spectra, 26 show these characteristics and are called YVLM sample (short for 'young very low mass')
in the following. For 11 the spectra are too 
noisy to make a decision, and the remaining 63 do not show these features. These 63 objects for which we can 
exclude that they are very low mass sources are listed in Appendix \ref{a1}.

\begin{figure}
\center
\includegraphics[width=8.5cm]{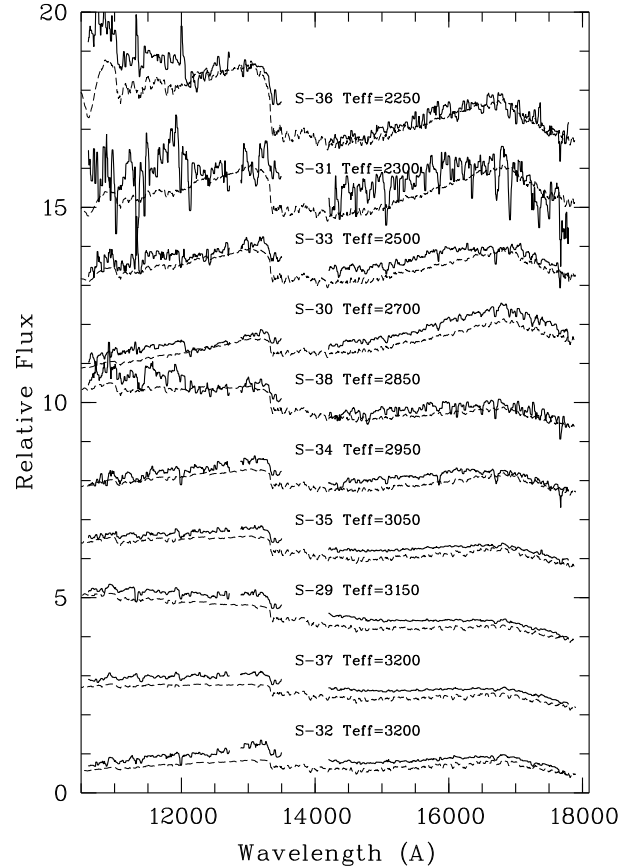} 
\caption{FMOS spectra for the 10 new brown dwarfs listed in Table \ref{t1}. The fluxes are on a
relative scale with arbitrary offsets. The observed spectrum is shown in black, the best fitting 
DUSTY model slightly offset, with thin, dashed lines. We masked out the Pa$\beta$ lines for clarity.
The identification numbers SONYC-NGC1333-X are abbreviated with S-X.
\label{f20}}
\end{figure}

From the 19 literature sources, 16 are re-identified and are part of the YVLM sample. The other 3 have spectra 
that are too noisy to identify the features. The remaining 10 YVLM objects are newly confirmed very low mass members of
NGC1333 and are listed in Table \ref{t1}. We use the nomenclature SONYC-NGC1333-X for these objects, where X is
a running number. Since we have listed objects 1-28 in \citet{2009ApJ...702..805S}, we continue here with no. 
29. Note that the list in \citet{2009ApJ...702..805S} contains some previously confirmed members. The spectra for
the 10 new objects are shown in Fig. \ref{f20}. 7 of the new objects come from our IZ catalogue, 
the remaining 3 from the JK plus Spitzer list (SONYC-NGC1333-31, 32, 33). 

\begin{deluxetable*}{cllccccccccl}
\tabletypesize{\scriptsize}
\tablecaption{New very low mass members in NGC1333. The IDs SONYC-NGC1333-X are abbreviated with S-X. 
\label{t1}}
\tablewidth{0pt}
\tablehead{\colhead{ID} & \colhead{$\alpha$(J2000)} & \colhead{$\delta$(J2000)} & 
\colhead{i' (mag)\tablenotemark{a}}  & \colhead{z' (mag)\tablenotemark{a}} & 
\colhead{J (mag)\tablenotemark{b}} & \colhead{K (mag)\tablenotemark{b}} &
\colhead{$A_V$\tablenotemark{c}} & \colhead{$A_V$\tablenotemark{d}} & 
\colhead{SpT\tablenotemark{e}} & \colhead{$T_{\mathrm{eff}}$\tablenotemark{f}} & 
\colhead{Other names\tablenotemark{g}}}
\tablecolumns{11}
\startdata
S-29 & 03 28 28.40 & +31 16 27.3 & 17.637 & 16.675 & 14.624  & 13.624  & 0 & 0 & M6.9      & 3150 &		  \\ 
S-30 & 03 28 31.08 & +31 17 04.1 & 23.650 & 21.235 & 16.823  & 14.079  & 9 & 9 & M9.3      & 2700 &		  \\ 
S-31 & 03 29 44.15 & +31 19 47.9 &        &        & 17.386  & 15.290  & 6 & 5 & $\sim$M9  & 2300 & Sp~132	  \\ 
S-32 & 03 29 03.21 & +31 25 45.2 &        &        & 15.798  & 13.832  & 4 & 4 & M7.1      & 3200 & MBO89, Sp~79  \\ 
S-33 & 03 29 03.95 & +31 23 30.8 &        &        & 17.143  & 14.932  & 7 & 5 & M8.3      & 2500 & MBO116, Sp~83 \\ 
S-34 & 03 29 06.94 & +31 29 57.1 & 20.839 & 19.225 & 16.541  & 14.774  & 4 & 4 & M7.0      & 2950 & Sp~90	  \\ 
S-35 & 03 29 10.18 & +31 27 16.0 & 18.988 & 17.711 & 15.547  & 14.127  & 2 & 2 & M7.4      & 3050 & MBO94, Sp~96  \\ 
S-36 & 03 29 25.84 & +31 16 41.8 & 23.392 & 21.432 & 18.53   & 17.07   & 2 & 0 & $\sim$L3  & 2250 &		  \\ 
S-37 & 03 29 30.54 & +31 27 28.0 & 17.356 & 16.344 & 13.808  & 12.598  & 1 & 2 & M6.9      & 3200 & MBO68	  \\ 
S-38 & 03 29 34.42 & +31 15 25.2 & 21.091 & 19.654 & 17.33   & 16.15   & 1 & 1 & M7.7      & 2850 &		  \\ 
\enddata
\tablenotetext{a}{i- and z-band photometry from \citet{2009ApJ...702..805S}}
\tablenotetext{b}{J- and K-band photometry from 2MASS or, for SONYC-NGC1333-36 and 38, from \citet{2009ApJ...702..805S}}
\tablenotetext{c}{Calculated from the J- and K-band magnitudes using Equ. 1}
\tablenotetext{d}{Corrected $A_V$ after spectral fitting, see Sect. \ref{s33}}
\tablenotetext{e}{Estimated using the HPI index as defined in Sect. \ref{s32}}
\tablenotetext{f}{Estimated by comparing the spectra to models, see Sect. \ref{s33}}
\tablenotetext{g}{identifiers are from \citet[][MBO]{2004AJ....127.1131W} and the Spitzer survey 
by \citet[][Sp]{2008ApJ...674..336G}}
\end{deluxetable*}

Many of our spectra show emission in Paschen $\beta$ at 1.28$\,\mu m$. If the emission originates from the
source itself, it would be a clear indication for ongoing accretion \citep{2004A&A...424..603N}. However, this 
is difficult to tell with fibre spectroscopy; our spectra may be affected by emission from the cloud material 
in NGC1333. Still, it is worth pointing out that the fraction of objects with Pa $\beta$ emission is 50\% in the 
YVLM sample and 32\% among the remaining objects. This indicates that the other sources might contain a number
of young stars in NGC1333 with temperatures $>3500$\,K.

In Fig. \ref{f4} we compare the FMOS spectra for two of our sources, both originally identified in 
\citet{2009ApJ...702..805S}, with spectra for young brown dwarfs from \citet[][left panel]{2007AJ....134..411M} 
and old dwarfs with similar spectral types \citep[][middle panel]{2008ApJ...681..579B,2006AJ....131.1007B}. 
At late M spectral types, the more rounded peak of the old objects becomes visible (compare the young
KPNO4 with the old LHS2924). However, for earlier types the difference between young and old is not
significant. The plot also illustrates that sufficient signal-to-noise is required to distinguish between
the two types of objects. Our spectra often do not have
the necessary quality to make this decision. From these arguments it follows that contamination by late 
M-dwarfs in the field cannot completely be excluded. Given the compact nature of the NGC1333 cluster 
(see Sect. \ref{s44}) and the low space densities of such objects \citep{2008A&A...488..181C}, the 
contamination is considered to be negligible ($\lesssim 1$).

\subsection{Spectral types}
\label{s32}

Spectral classification for cool dwarfs in the near-infrared relies mostly on the broad H$_2$O absorption features 
mentioned above. A number of spectral indices has been suggested in the literature. Relevant for us
are the indices that are calibrated for young brown dwarfs and thus take into account the fact that the 
absorption features are sensitive to gravity. Such indices have been proposed by \citet{2004AJ....127.1131W},
\citet{2007ApJ...657..511A}, and \citet{2009MNRAS.392..817W}. We find, however, that these schemes are of
limited use for our low-resolution and (mostly) low signal-to-noise spectra, mainly because the wavelength offset
between the two intervals from which the index is derived is relatively small, e.g. 0.058$\,\mu m$ for the H$_2$O index
\citep{2007ApJ...657..511A} or 0.103$\,\mu m$ for the WH index \citep{2009MNRAS.392..817W}.

For our purposes we require a flux ratio that maximises the contrast between the numerator and the denominator. 
This is achieved by using the flux ratio between 1.675-1.685$\,\mu m$ and 1.495-1.505$\,\mu m$. The first interval is 
chosen because it corresponds to the position of the H-band peak for young objects with M6-M9 spectral type. The 
second interval marks the lowest flux level on the blue side of the H-band peak for such objects.

\begin{deluxetable}{lccccc}
\tabletypesize{\scriptsize}
\tablecaption{Objects used to calibrate the H-peak index. 
\label{t3}}
\tablewidth{0pt}
\tablehead{
\colhead{Name\tablenotemark{a}} & \colhead{SpT\tablenotemark{b}}  & \colhead{J (mag)\tablenotemark{c}} 
& \colhead{K (mag)\tablenotemark{c}} & \colhead{$A_V$ (mag)\tablenotemark{d}} & \colhead{HPI\tablenotemark{d}}}
\tablecolumns{6}
\startdata
ITG2	& M7.25  &  11.540 & 10.097 & 2.40  & 1.050  \\
J0444	& M7.25  &  12.195 & 10.761 & 2.35  & 1.124  \\
CFHT6	& M7.5   &  12.646 & 11.368 & 1.51  & 1.135  \\
KPNO2	& M7.5   &  13.925 & 12.753 & 0.93  & 1.136  \\
KPNO5	& M7.5   &  12.640 & 11.536 & 0.56  & 1.137  \\
CFGH3	& M7.75  &  13.724 & 12.367 & 1.94  & 1.184  \\
J0441	& M7.75  &  13.730 & 12.220 & 2.77  & 1.098  \\
CFHT4	& M7.0   &  12.168 & 10.332 & 4.53  & 1.009  \\
MHO4	& M7.0   &  11.653 & 10.567 & 0.47  & 1.174  \\
KPNO7	& M8.25  &  14.521 & 13.271 & 1.37  & 1.121  \\
KPNO1	& M8.5   &  15.101 & 13.772 & 1.78  & 1.203  \\
KPNO6	& M8.5   &  14.995 & 13.689 & 1.63  & 1.185  \\
KPNO9	& M8.5   &  15.497 & 14.185 & 1.69  & 1.155  \\
LRL405  & M8.0   &  15.28  & 13.91  & 2.01  & 1.126  \\ 
J0457	& M9.25  &  15.771 & 14.484 & 1.56  & 1.240  \\
KPNO4	& M9.5   &  14.997 & 13.281 & 3.91  & 1.342  \\
KPNO12  & M9.0   &  16.305 & 14.927 & 2.01  & 1.261  \\
J1207	& M8.0   &  12.995 & 11.945 & 0.81  & 1.226  \\ 
J1139	& M9.0   &  12.686 & 11.503 & 0.99  & 1.336  \\
J1245	& M9.5   &  14.518 & 13.369 & 0.81  & 1.306  
\enddata
\tablenotetext{a}{see SpeX Prism Spectral Libraries:\\ {\tt http://pono.ucsd.edu/\textasciitilde adam/browndwarfs/spexprism/}}
\tablenotetext{b}{spectral types from \citet{2007AJ....134..411M} for the first 17 and 
\citet{2007ApJ...669L..97L} for the remaining 3}
\tablenotetext{c}{photometry from 2MASS, except LRL405 \citep{2003A&A...409..147P,2003ApJ...593.1093L}}
\tablenotetext{d}{derived as described in the text}
\end{deluxetable}

\begin{figure*}
\center
\includegraphics[width=5.9cm]{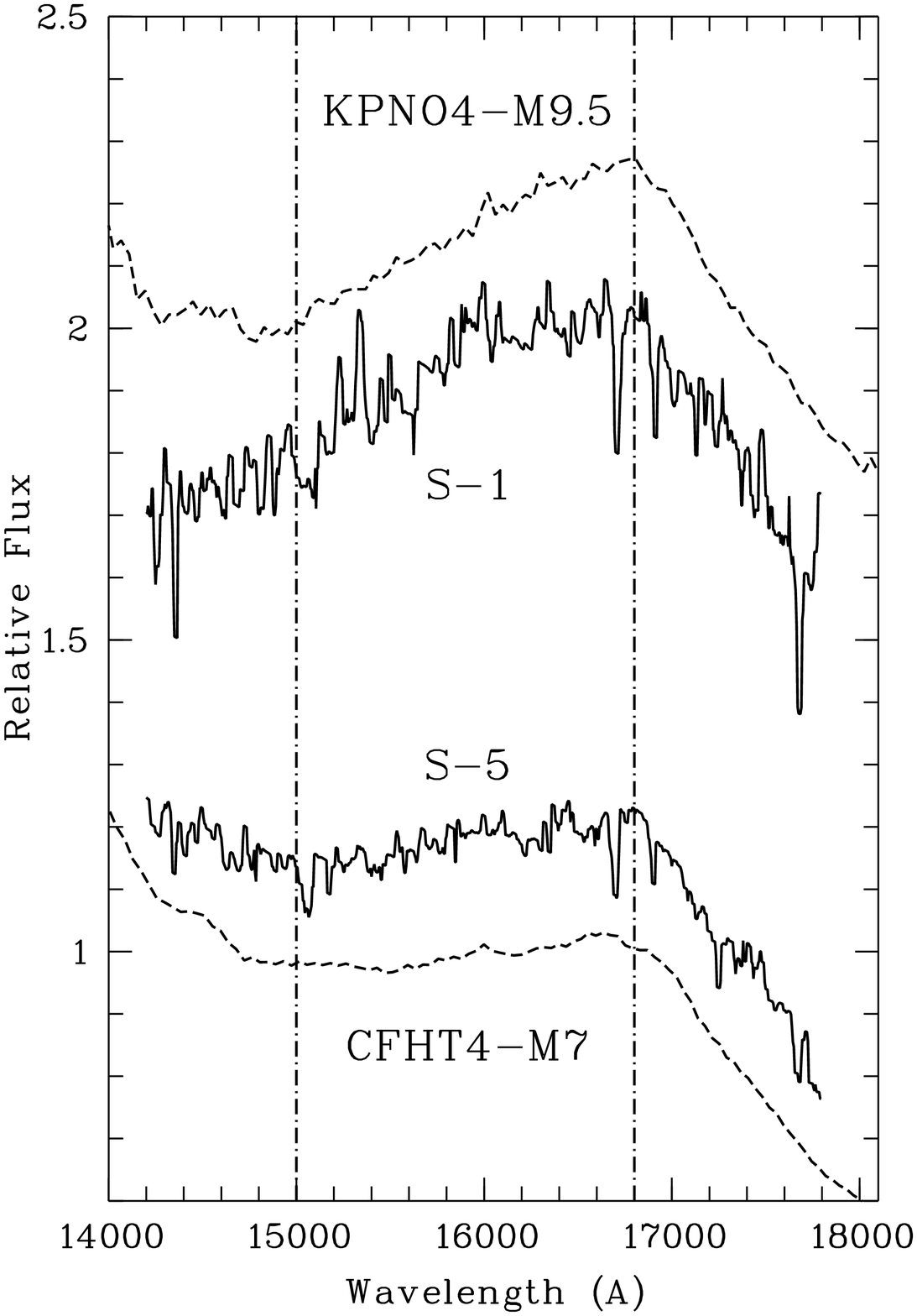} 
\includegraphics[width=5.9cm]{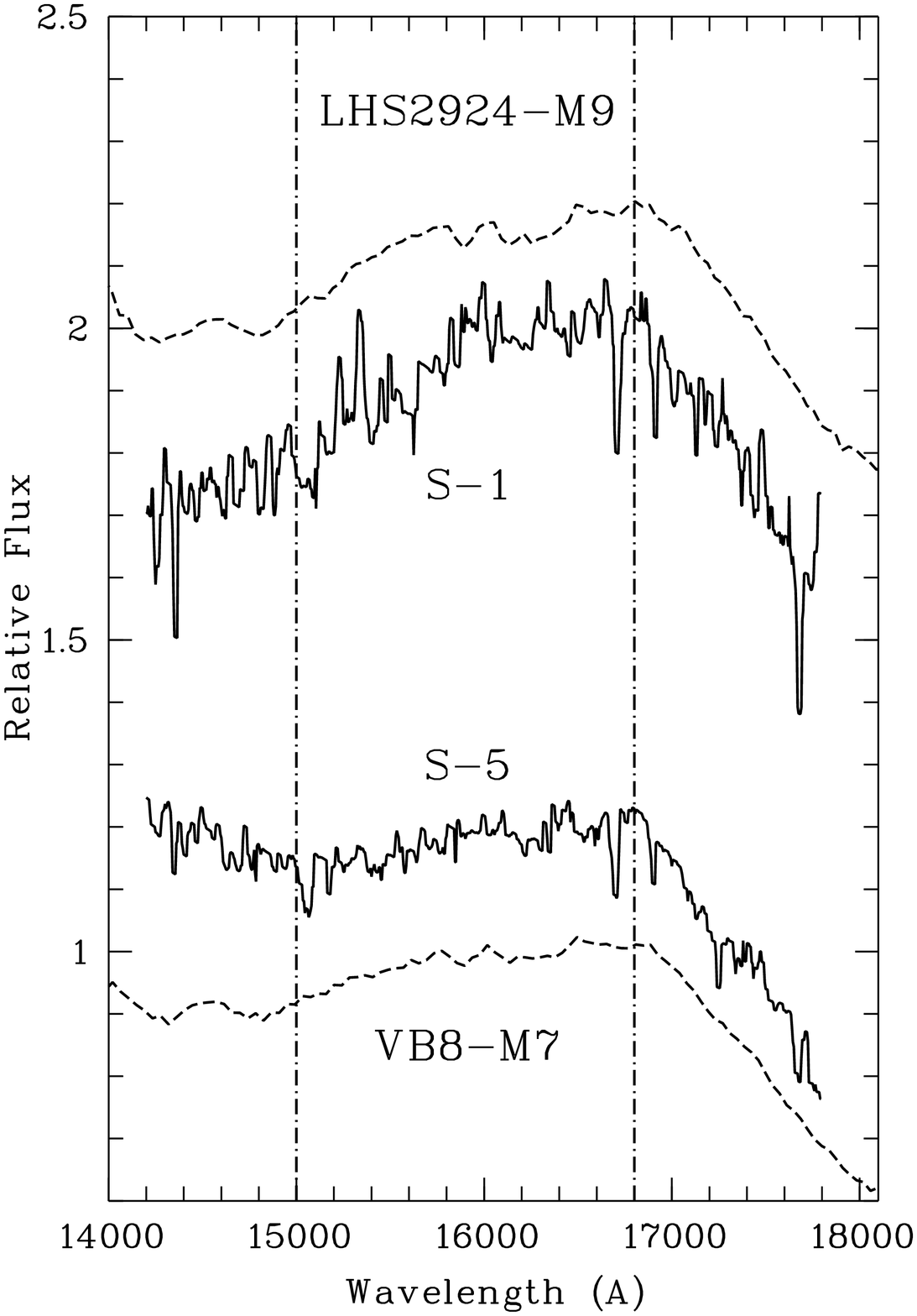} 
\includegraphics[width=5.9cm]{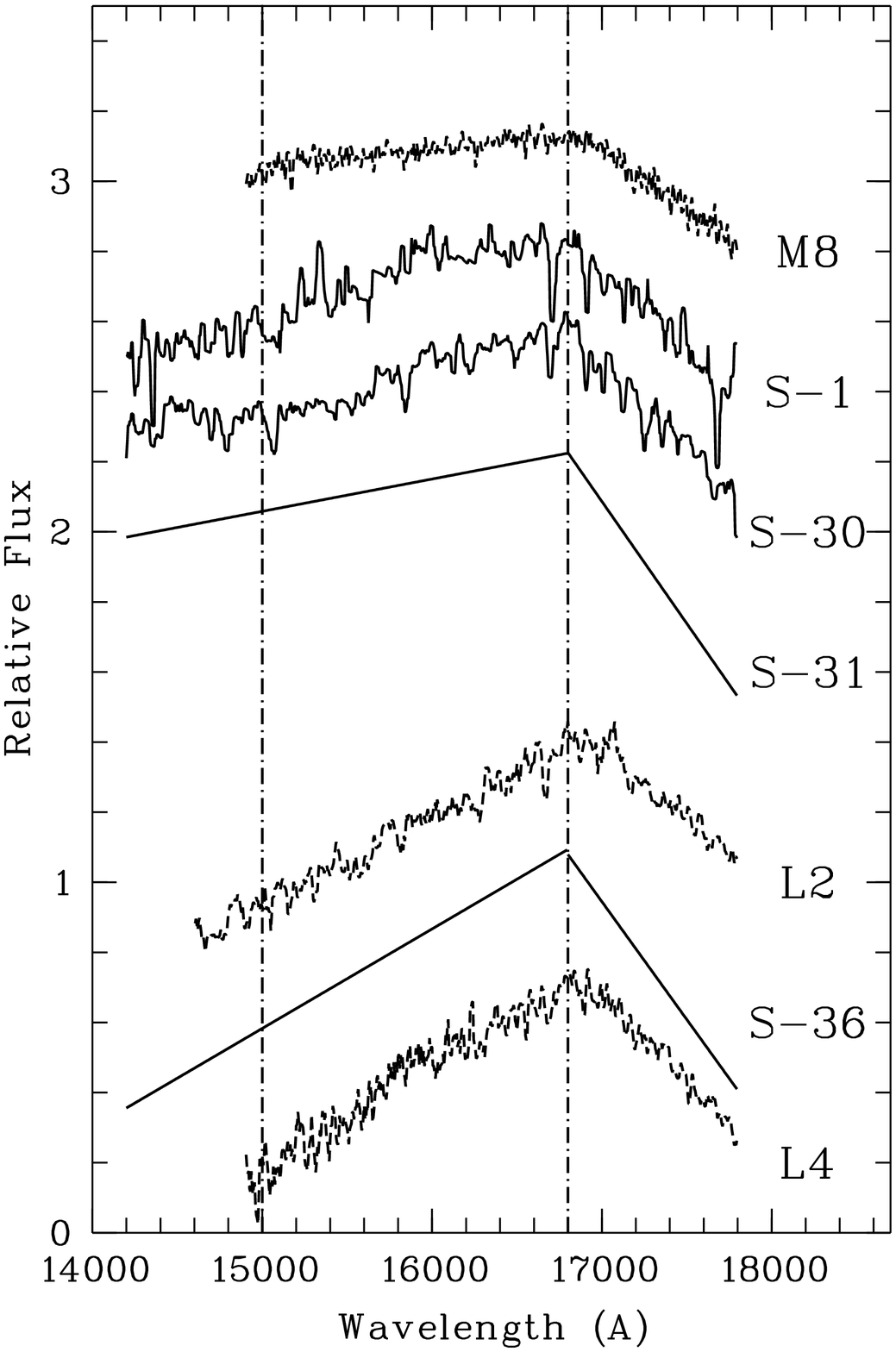}
\caption{Dereddened spectra of SONYC objects in NGC1333 (solid lines) compared with different comparison 
spectra (dashed lines). Spectra are offset by a constant value for clarity. The spectral range for the HPI 
is indicated with dashed lines. The SONYC-NGC1333-X identification numbers are abbreviated with S-X in all 
three panels. {\bf Left panel:} Young brown dwarfs in Taurus \citep{2007AJ....134..411M}. 
{\bf Middle panel:} M dwarf spectral standards in the field \citep{2008ApJ...681..579B,2006AJ....131.1007B}.
{\bf Right panel:} Young substellar companions (M8 - HIP78530B \citep{2011ApJ...730...42L}, 
L2 - 2M1207B \citep{2010A&A...517A..76P}, L4 - 1RXSJ1609-2105B \citep{2010ApJ...719..497L}.
In the right panel, we approximate the spectral slope for two of the faintest objects with a 
linear fit, the full spectra are shown in Fig. \ref{f20}.
\label{f4}}
\end{figure*}

This index, dubbed HPI for H-peak index, is illustrated in the left and middle panel of Fig. \ref{f4} where we 
show literature spectra for young and old brown dwarfs in Taurus with spectral types M7 and M9.5 compared with 
FMOS data for two of the brown dwarfs in NGC1333. The H-band peak clearly is highly sensitive to the spectral 
type in this regime. We expect that this index increases from mid M to late M spectral types. The figure also 
demonstrates the advantages of this index at low signal-to-noise ratio.

To calibrate the HPI, we use literature spectra for 20 young brown dwarfs with spectral types M7-M9.5, for which
spectra are publicly available in the SpeX Prism Spectral Libraries \citep{2007AJ....134..411M,2007ApJ...669L..97L}. 
Their spectra have been dereddened using the reddening law $A_{\lambda} = (\lambda / 1.235\,\mu m)^{-1.61} \cdot A_J$ 
and the relation $A_J = 0.3088 \cdot A_V$. The optical extinction $A_V$ is derived from their $J-K$ colours: 
\begin{equation}
A_V = [(J-K) - (J-K)_0] / 0.1844
\end{equation}
These relations are based on the extinction law from \citet{1989ApJ...345..245C} for $R_V = 4.0$, which is used 
throughout this paper. We note that $R_V$ varies within star forming regions typically from 3 to 5; the adopted 
value is an average from the values measured by \citet{1989ApJ...345..245C} for diffuse and dense regions of the 
interstellar medium. It is also a reasonable average for our target region NGC1333 \citep{1990Ap&SS.166..315C}.

For the reasons outlined in \citet{2009ApJ...702..805S} we use $(J-K)_0 = 1.0$. The resulting uncertainty in $A_V$ is 
$\sim 1$\,mag. For example it is possible that we overestimate $A_V$ due to the presence of K-band 
excess from a disk. This induces an uncertainty of up to 0.04 in the HPI. Note that the literature spectral types 
for the calibrators have been determined in the optical by comparison with templates \citep[e.g.][]{2002ApJ...580..317B}. 
The internal accuracy of these optical types is typically $\pm 0.25$ subtypes. The calibrators are listed in
Table \ref{t3}.

\begin{figure}
\center
\includegraphics[width=6.0cm,angle=-90]{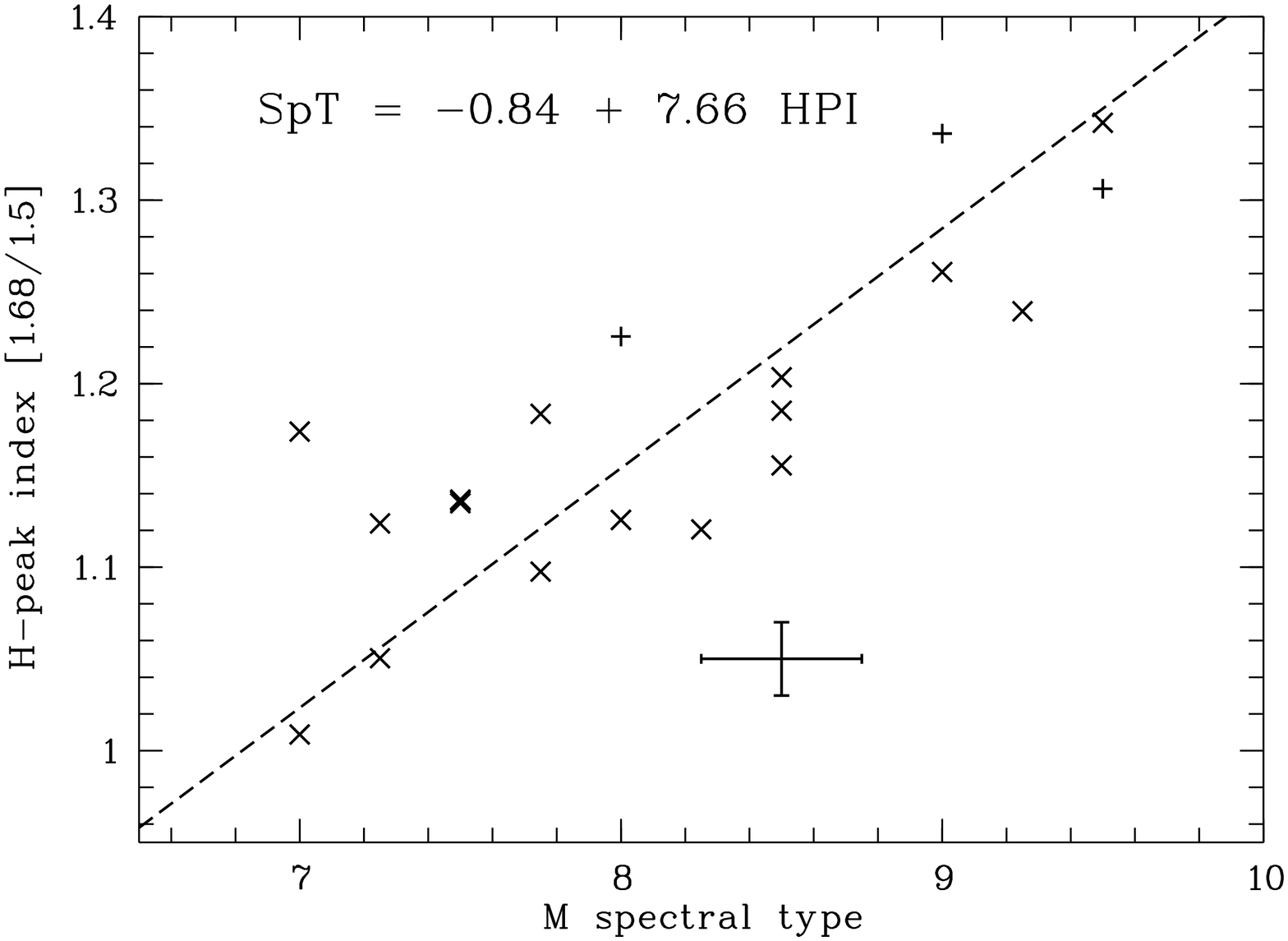} 
\caption{H-peak index (HPI) calculated as flux ratio between 1.675-1.685$\,\mu m$ and 1.495-1.505$\,\mu m$ 
plotted vs. (optical) spectral type for published spectra from \citet[crosses,][]{2007AJ....134..411M} 
and \citet[plusses,][]{2007ApJ...669L..97L}. The correlation after excluding the outlier MHO4 at M7 is
$\mathrm{SpT} = -0.84 + 7.66\cdot \mathrm{HPI}$ (dashed line). A typical errorbar is overplotted.
\label{f3}}
\end{figure}

In Fig. \ref{f3} we plot the HPI for the 20 calibrators. The plot shows the expected 
correlation between index and spectral type. The one outlier at spectral type M7 corresponds to 
the object MHO4 and deviates in its HPI by $3\sigma$ from a linear fit. Nothing particular can 
be seen in its spectrum which would explain this discrepancy. Based on the H-band peak, 
MHO4 appears to have a later spectral type than indicated in the literature.
Excluding MHO4, we derive a correlation of 
\begin{equation}
\mathrm{SpT} = -0.84 + 7.66\cdot \mathrm{HPI}
\end{equation}
with an 1$\sigma$ scatter of 0.37 subclasses. The 1$\sigma$ scatter in HPI around the correlation is 0.039. As 
outlined above, this can be attributed to the uncertainty in the extinction. The HPI is properly calibrated
for types M7-M9.5, but may also hold for later spectral types as the H-band peak increases in strength in the
L-type regime (see below).


We note that this correlation does not hold for field dwarfs. As seen in the middle panel of Fig. \ref{f4} 
these objects have flatter H-band peaks which results in lower values for the HPI for a given spectral type. 

Using this index we determined spectral types for all 26 objects in the YVLM sample, after dereddening
using the corrected extinction derived in Sect. \ref{s33}. 9 of them have published spectral types. Five of 
them (MBO69, MBO54, MBO77, ASR29, ASR83) are based on an index defined in the K-band 
\citep{2004AJ....127.1131W,2007AJ....133.1321G}. The other four have the Spitzer IDs 131, 104, 118, and 46 in 
\citet{2009AJ....137.4777W} and were classified in the optical. Excluding MBO54, ASR29, Sp~104 
and Sp~118 which have a published spectral type of M6 or earlier, for which the HPI is not properly calibrated, 
the HPI types deviate by -0.4, -0.2, -0.2, -0.1, +0.6 subtypes from the published ones, which provides some 
reassurance in the usefulness of the HPI. The 10 new objects have spectral types of M6.9 or later,
classifying them as likely brown dwarfs. The spectral types for the new and known sources are listed in 
Table \ref{t1} and \ref{t2} respectively.


In the right panel of Fig. \ref{f4} we compare FMOS data for the four latest-type SONYC objects with 
spectra for young ultracool objects with published spectral types: the M8 companion to HIP78530 
\citep{2011ApJ...730...42L},  the L2 companion to the TW Hya brown dwarf 2M1207 \citep{2010A&A...517A..76P}, 
and the L4 companion to 1RXSJ1609-2105B \citep{2010ApJ...719..497L}, all three with ages of 5-10\,Myr. 
Using the HPI, we get spectral types of M8, L2, and L3.5 for these three comparison objects, consistent with 
the literature values, which indicates that HPI could be useful for spectral typing of young early L-dwarfs
as well. In this plot we approximate the spectral slopes for the two faintest objects in our sample with linear 
fits on either side of the H-band peak, to facilitate the comparison. The plot demonstrates that objects 
SONYC-NGC1333-1, 30, 31 are about or later than M8 and clearly earlier than L2, in line with our classification. 
The object SONYC-NGC1333-36 compares well with the L2 and L4 templates, which makes it the coolest object 
discovered in NGC1333 so far. 

We note that the redward slope
of the H-band peak appears anomalously steep for SONYC-NGC1333-31 and 36. This is not introduced by the linear fit or the 
treatment of the spectra, and it is not seen in the other FMOS spectra for NGC1333 or $\rho$\,Oph (Muzic et al., subm.), 
which makes a calibration problem unlikely. The effect is difficult to explain, as these two spectra are very noisy, 
but it could in principle be a real feature, especially since the data is well-matched by the model spectrum
(see Sect. \ref{s33}). Better quality spectra are needed to verify the parameters for these objects. Since the HPI 
is defined for the blueward slope of the H-band peak, this does not affect the spectral typing.

\subsection{Model fitting}
\label{s33}

For the 26 objects in the YVLM sample the FMOS spectra were compared with AMES-DUSTY models 
\citep{2001ApJ...556..357A} for low gravity ($\log{g} = 3.5$ or, if not available, $4.0$) low-mass 
stars\footnote{downloaded \tt{from\\ http://phoenix.ens-lyon.fr/Grids/AMES-Dusty/SPECTRA/}}. Using the 
extinction law described in Sect. \ref{s32} we calculated a model grid for $T_{\mathrm{eff}} = 1500$ to 
3900\,K in steps of 100\,K and $A_V = 1$ to 20\,mag in steps of 1\,mag. The models were binned to 
5\,\AA, the same binsize as the FMOS data. 

The fitting was done in a semi-interactive way. Since extinction and effective temperature cannot 
be determined separately with low-resolution, low signal-to-noise spectra, we started by adopting the $A_V$
determined from the $J-K$ colour, using Equ. (1). For each of the YVLM objects, we calculated the following
test quantity ($O$ - observed flux; $M$ - model flux; $N$ - number of datapoints):
\begin{equation}
\chi = \frac{1}{N} \sum_{i=1}^{N} \frac{(O - M)^2}{M}
\end{equation} 
This was done for the series of models using the 'photometric' $A_V$; we selected the one with the minimum 
$\chi$, which is typically between 0.005 and 0.05 (with one exception with 0.2). This means that the 
average deviation between observed and model spectrum in a given wavelength bin of 5\,\AA~is in the same
range as the noise in the observations.

Usually a few model spectra (2-4) give indistinguishable $\chi$; we adopt the average $T_{\mathrm{eff}}$
from these best fitting models. A visual inspection of the observed spectra with models for a range of 
temperatures shows that clear discrepancies are visible for temperatures which are $>200$\,K different 
from the adopted value, i.e. the uncertainty in the adopted values is $\pm 200$\,K. We note that effective
temperatures are necessarily model dependent; our values should only be interpreted in the context of 
the AMES-DUSTY models. The best fitting models are plotted as thin, dashed lines in Fig. \ref{f20} for the 
10 newly identified objects.

In 6/10 cases this initial fit is already convincing. In two more cases, it can be improved by slightly 
adjusting $A_V$ by 1\,mag, which is within the uncertainty for $A_V$ (see Sect. \ref{s32}). The two 
remaining cases give the best fit when $A_V$ is changed by 2\,mag compared with the initial estimate.
In Table \ref{t1} we list the photometric and adjusted value for $A_V$ and the best fit value for the 
effective temperature. 

For SONYC-NGC1333-36, the coolest object in our sample, we find an effective temperature of 2250\,K, which is
significantly  higher than the published values for the two comparison objects shown in Fig. \ref{f4} (right panel).
For 1RXSJ1609-2105B \citet{2010ApJ...719..497L} find $1800\pm 200$\,K based on DUSTY and Drift-Phoenix
\citep{2008ApJ...675L.105H} models. For 2M1207B, two independent groups determined $1600\pm 100$\,K, again 
based on comparison with DUSTY spectra \citep{2010A&A...517A..76P,2007ApJ...657.1064M}, although 
\citet{2011ApJ...732..107S} suggest that the actual value might be as low as 1000\,K. The DUSTY spectra 
for $<2000$\,K are clearly not in agreement with our spectrum for SONYC-NGC1333-36. 
One possible explanation is that model fitting for the comparison objects has been done over the full 
near-infrared range, whereas in our case $T_{\mathrm{eff}}$ is mostly fixed by the shape of the H-band 
peak. We will revisit this issue in Sect. \ref{s34}.

\begin{figure}
\center
\includegraphics[width=6.0cm,angle=-90]{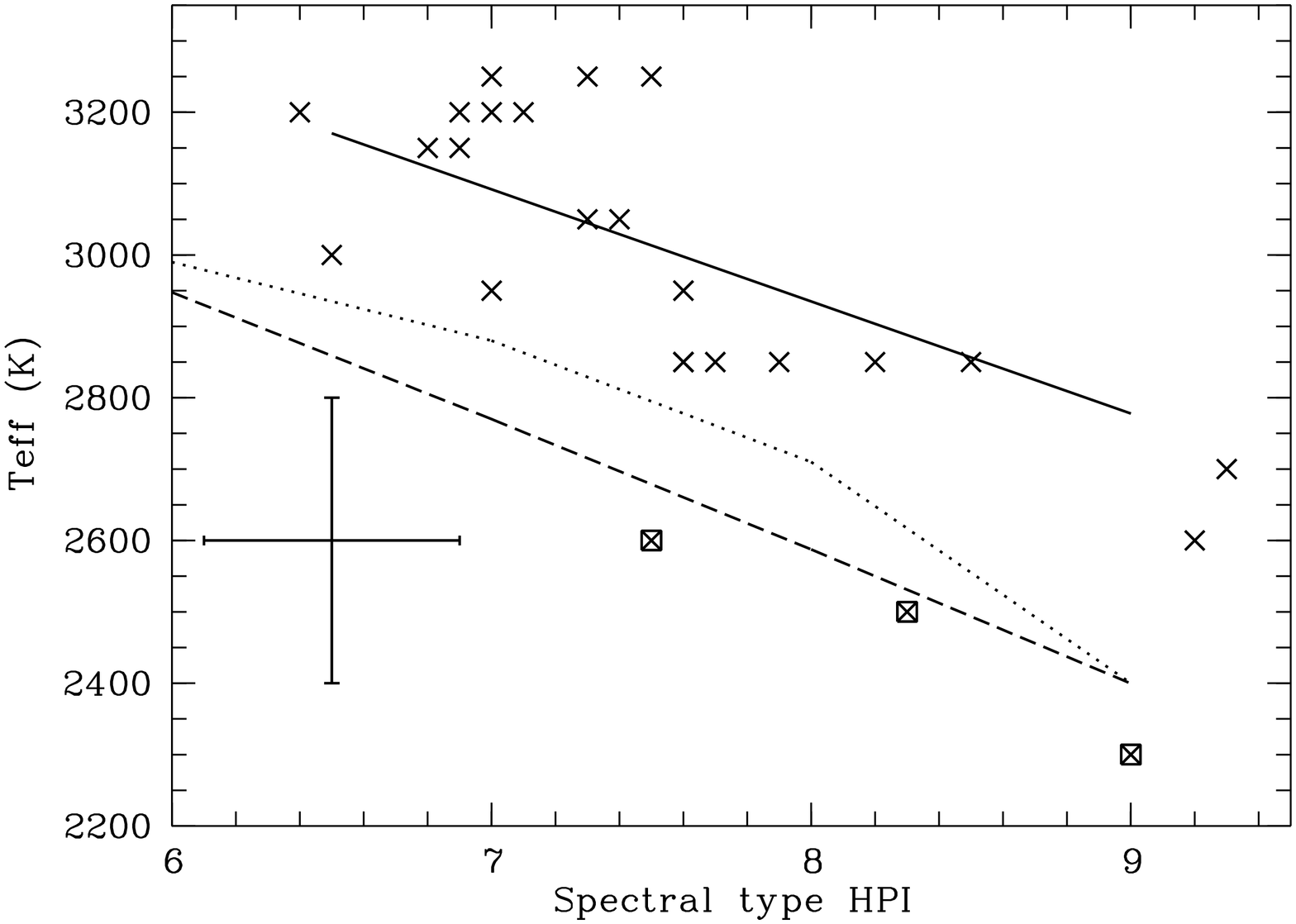} 
\caption{Comparison between the spectral types and effective temperatures for the YVLM sample, as 
determined in this paper (Sect. \ref{s32} and \ref{s33}). A typical errorbar is overplotted. The solid 
line is a linear fit to the datapoints excluding the 3 outliers marked with squares. We also show the 
effective temperature scales by \citet[][dashed line]{2008ApJ...689.1127M} 
and \citet[][dotted line]{2003ApJ...593.1093L}. 
\label{f12}}
\end{figure}

In Fig. \ref{f12} we compare the derived effective temperatures with the spectral types determined
with the HPI in Sect. \ref{s32}. The expected trend -- cooler temperatures correspond to later 
spectral types -- is visible, with only three exceptions. For two of these exceptions (SONYC-NGC1333-17 and
-31), the signal-to-noise ratio in the spectrum is very poor ($\sim 10$), i.e. the uncertainties in spectral 
type and temperature are large. For SONYC-NGC1333-17 there are two published spectral types of M8 and M8.7,
which would move the datapoint closer to the general trend. The third outlier (SONYC-NGC1333-33) has an 
unusual dip in the spectrum around 1.65$\,\mu m$, which decreases our spectral type estimate by $\sim 1$
subtype, but does not affect the model fitting. Excluding these three outliers, the datapoints are
well-approximated by a linear trend (solid line): $T_{\mathrm{eff}} = 4191 - 157 \times \mathrm{SpT}$ 
(where SpT corresponds to the M subtype).

The plot also shows the effective temperature scales by 
\citet[][dashed line]{2008ApJ...689.1127M} and \citet[][dotted line]{2003ApJ...593.1093L}, which
are derived using the optical portion of the spectrum. The two scales agree fairly well with each other, 
although the Mentuch et al. scale is an extrapolation and has not been directly calibrated for 
$T_{\mathrm{eff}} <3000$\,K. The trend seen in our data is consistent with these lines. Our datapoints 
are, however, mostly above the two lines, indicating that we systematically overestimate the temperatures 
(by $\sim 200$\,K) or the spectral types (by 0.5-1 subtypes). 

We explored possible reasons for this discrepancy. One option is a systematic error in the extinction.
Decreasing $A_V$ by 1\,mag for all objects would shift their spectral types to earlier types, but only by 
$\sim 0.3$ subtypes, not sufficient to explain the effect. Moreover, this would also increase the best estimate 
for $T_{\mathrm{eff}}$ by typically 200\,K and thus cause larger disagreement between datapoints and published 
effective temperature scales. The inverse is true as well -- increasing $A_V$ would lead to lower 
$T_{\mathrm{eff}}$, as required, but also to later spectral types. Thus, systematic changes in $A_V$ do not 
resolve the problem. Given the good agreement between our spectral types and literature values (Sect. \ref{s32}), 
we suspect that the offset is most likely due to a problem with the effective temperatures and could indicate
issues with the used model spectra. A more extensive comparison of the effective temperature scales is 
beyond the scope of this paper.

\subsection{Spectral flux distributions}
\label{s34}

\begin{figure*}
\center
\includegraphics[width=4.0cm,angle=-90]{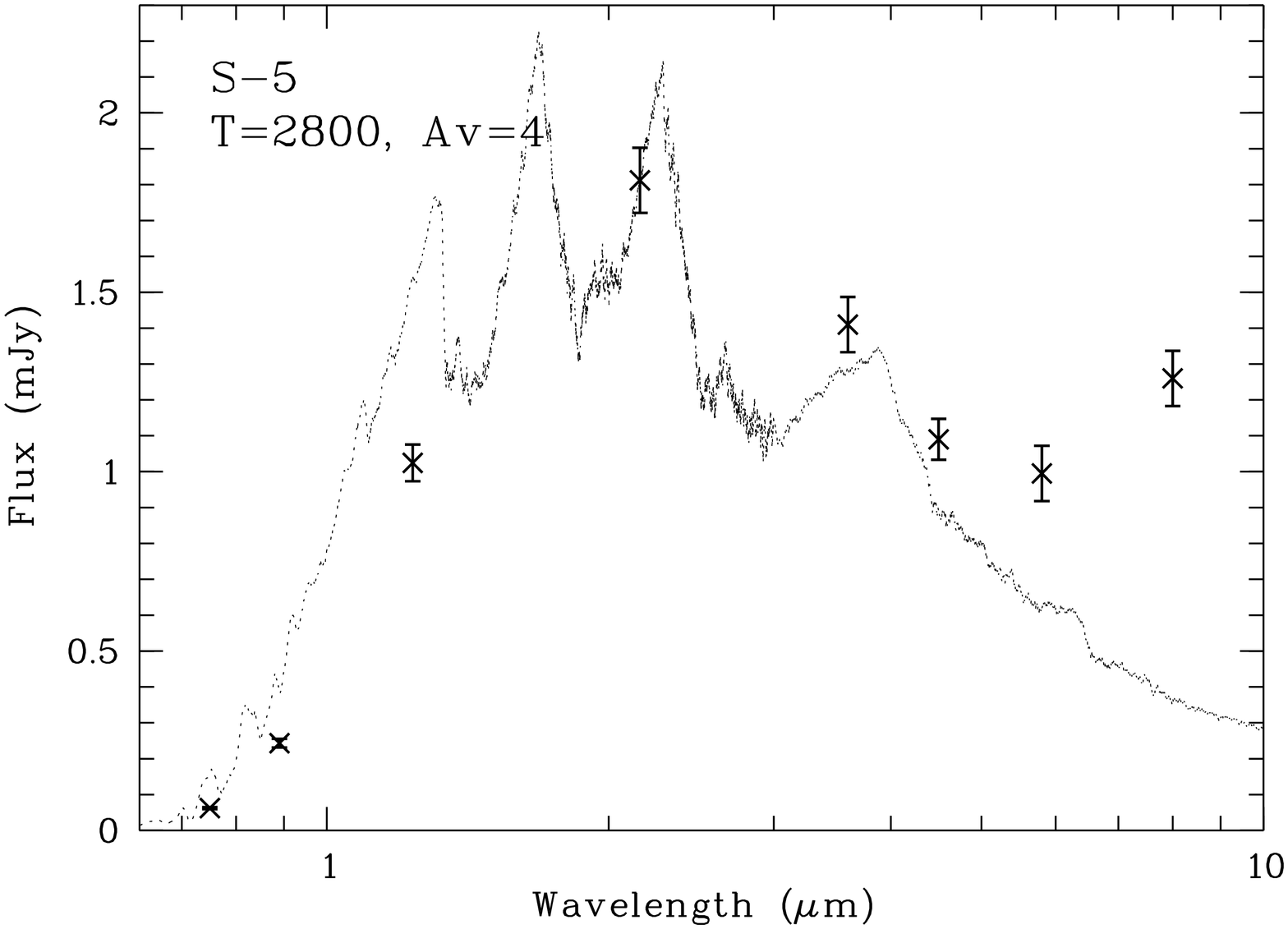} 
\includegraphics[width=4.0cm,angle=-90]{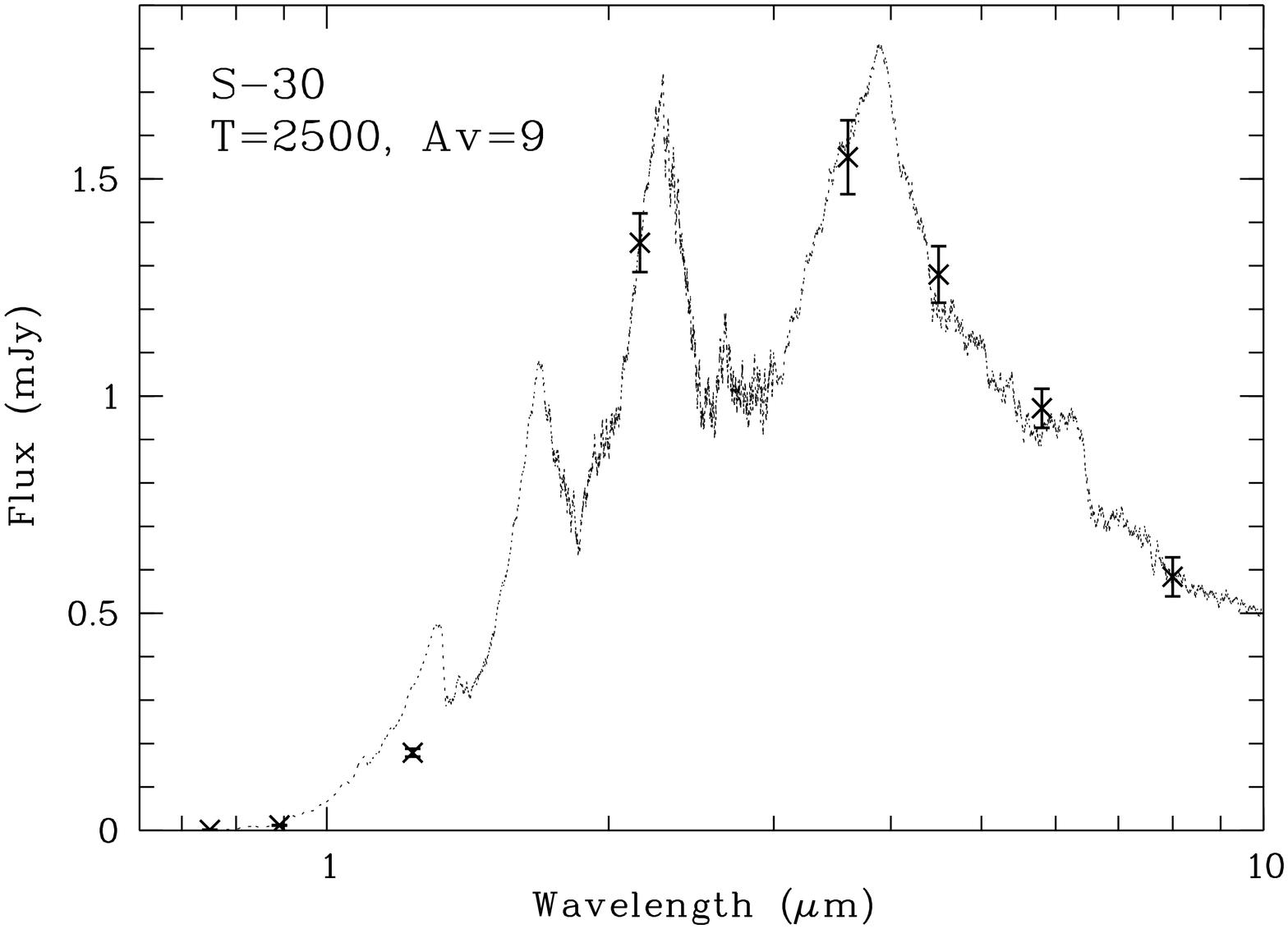} 
\includegraphics[width=4.0cm,angle=-90]{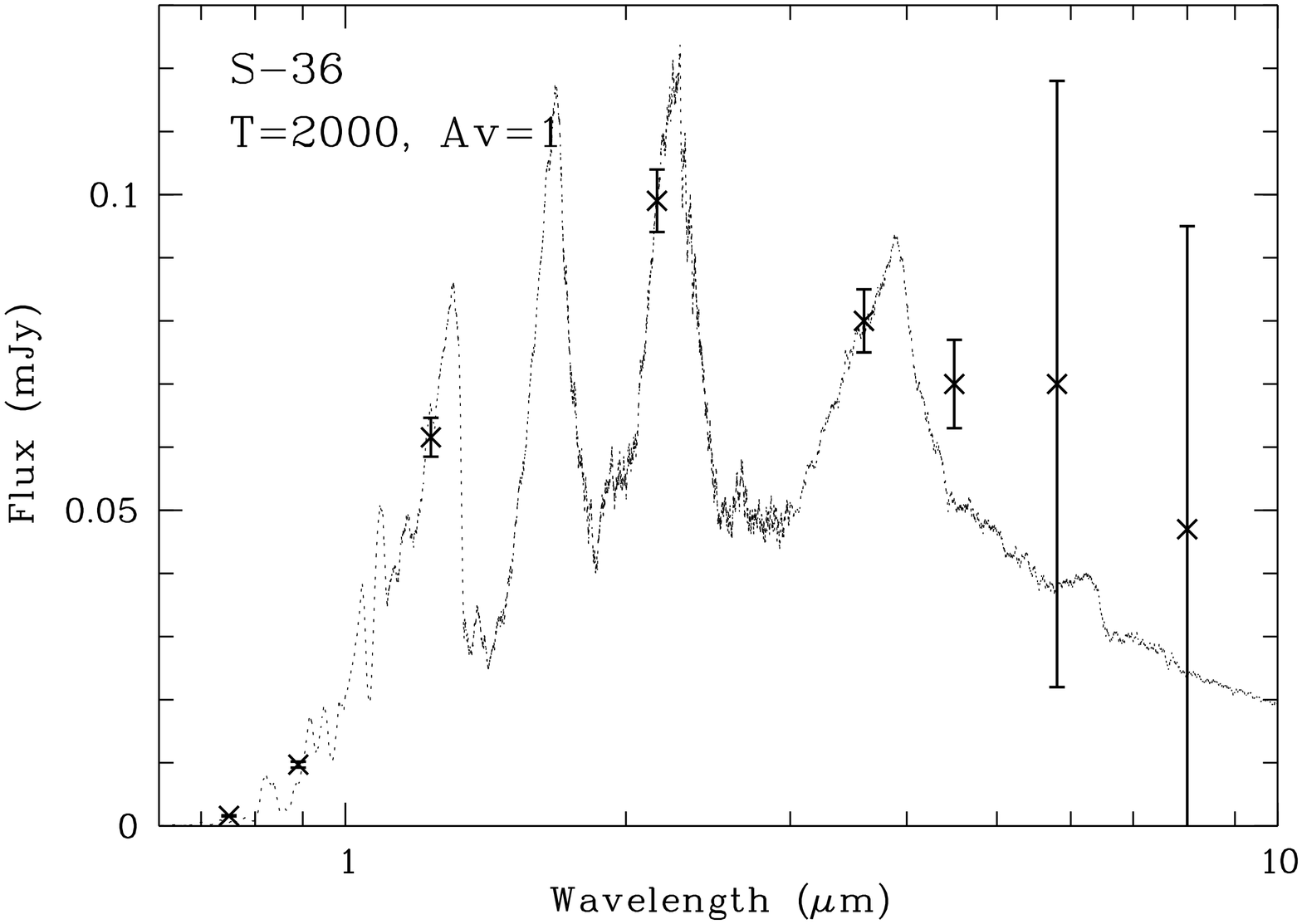} 
\caption{Photometric spectral flux distributions for three of the coolest members in NGC1333 in comparison with
AMES-DUSTY model spectra. The disk excess for SONYC-NGC1333-5 is clearly visible. The SONYC-NGC1333-X
identification numbers are abbreviated with S-X in all three panels. Errorbars for the fluxes are 
overplotted, but they are too small to be visible in some of the blue bands.
\label{f5}}
\end{figure*}

To further test the spectral fitting from Sect. \ref{s33} we compare the full photometric spectral flux
distributions (SFD) for some selected sources, as far as available, with the AMES-DUSTY model spectra. We include
our own photometry in izJK as well as Spitzer/IRAC data from the C2D 'HREL' catalogue \citep{2009ApJS..181..321E}. 
In Fig. \ref{f5} we show 3 examples. 

Object SONYC-NGC1333-5 is well-matched
by a model spectrum for $T_{\mathrm{eff}}$ of 2800-2900\,K for $A_V=4$\,mag, in line with the parameters derived
from the FMOS spectrum (Table \ref{t2}). This object shows colour excess redwards of 5$\,\mu m$, presumably
due to the presence of a disk (see Sect. \ref{s45}). The best match for object SONYC-NGC1333-30 is obtained for $T_{\mathrm{eff}}$ of 
2500-2700 with $A_V$ of 9-10\,mag, i.e. the object may be slightly cooler than listed in Table \ref{t1}, but still
within the uncertainty. For the coolest object SONYC-NGC1333-36 a good match is found for temperatures between 1900 and 2100\,K
and $A_V$ of 0-2\,mag. This is again somewhat cooler than the estimate given in Table \ref{t1}. 

In general, when comparing with the full SFD the results are similar to the spectral fits to individual 
wavelength bands in the near-infrared, maybe except for the regime below 2500\,K where the SFD comparison yields
lower temperatures by $\sim 200$\,K. The comparison also illustrates that the ideal dataset for a characterisation
of young brown dwarfs would be a spectrum covering the entire near-infrared domain from 1 to 8$\,\mu m$, thus including 
five broadband features. 

\section{The brown dwarf population in NGC1333}
\label{s4}

The newly identified very low mass objects in this paper add to the substantial number in the NGC1333 region that 
have already been confirmed in the literature. In Table \ref{t2} we compile all previously spectroscopically 
confirmed objects with spectral types of M5 or later or effective temperatures of 3200\,K or below, from  
\citet{2009ApJ...702..805S}, \citet{2004AJ....127.1131W}, \citet{2007AJ....133.1321G}, and \citet{2009AJ....137.4777W}. 
Whenever possible, we also re-measured the HPI spectral types for the sources identified in \citet{2009ApJ...702..805S}, 
based on their MOIRCS spectra. In the Table, we list coordinates, photometry, spectral types, effective temperatures, 
and alternative names. Adding the 10 objects discovered in this paper, the entire sample comprises 51 objects.

Not all these objects are brown dwarfs; some are very low mass stars. Based on the COND, DUSTY, and BCAH 
isochrones \citep{2003A&A...402..701B,2000ApJ...542..464C,1998A&A...337..403B}\footnote{downloaded from\\
{\tt http://perso.ens-lyon.fr/isabelle.baraffe/}}, the hydrogen burning 
limit at 1\,Myr would be reached at effective temperatures of 2800-2900\,K, which is found to correspond 
to an (optical) spectral type of M6-M7 \citep{2003ApJ...593.1093L,2008ApJ...689.1127M}. In Table \ref{t2}
we made the cut at M5 to include all borderline cases as well. Taking this into account, the number of 
confirmed brown dwarfs in NGC1333 is about $35 \pm 5$. This is currently one of the largest and best 
characterised populations of substellar objects in a single star forming region. In the following we 
will investigate the mass function, spatial distribution, and disk properties for this sample.

\begin{deluxetable*}{lllcclll}
\tabletypesize{\scriptsize}
\tablecaption{Previously confirmed very low mass members in NGC1333. The IDs SONYC-NGC1333-X 
are abbreviated with S-X. \label{t2}}
\tablewidth{0pt}
\tablehead{
\colhead{ID} & \colhead{$\alpha$(J2000)} & \colhead{$\delta$(J2000)} & \colhead{J (mag)\tablenotemark{1}} & 
\colhead{K (mag)\tablenotemark{1}} &
\colhead{SpT\tablenotemark{2}} & \colhead{$T_{\mathrm{eff}}$\tablenotemark{2}} & \colhead{Other names\tablenotemark{3}}} 
\tablecolumns{8}
\startdata
S-1    &  03 28 47.66 & +31 21 54.6 & 17.55  & 15.24  & M9.2$^a$                                & 2600$^a$, 2800$^b$ & MBO139              \\
S-2    &  03 28 54.92 & +31 15 29.0 & 15.990 & 14.219 & M7.9$^a$, M6.5$^b$, M8$^d$, M8.6$^f$    & 2850$^a$, 2600$^b$ & ASR109, Sp~60       \\ 
S-3    &  03 28 55.24 & +31 17 35.4 & 15.090 & 13.433 & M7.9$^c$, M8.2$^f$                      & 2900$^b$           & ASR38               \\
S-4    &  03 28 56.50 & +31 16 03.1 & 18.17  & 16.73  & M9.6$^f$                                & 2500$^b$  	     &                     \\
S-5    &  03 28 56.94 & +31 20 48.7 & 15.362 & 13.815 & M7.6$^a$, M6$^b$, M6.8$^c$              & 2850$^a$, 2900$^b$ & MBO91, Sp~66        \\ 
S-6    &  03 28 57.11 & +31 19 12.0 & 17.24  & 15.34  & M7.3$^a$, M8$^b$, M8.0$^f$              & 3250$^a$, 2700$^b$ & MBO148, ASR64, Sp~23\\ 
S-7    &  03 28 58.42 & +31 22 56.7 & 15.399 & 13.685 & M6.5$^b$, M7.1$^c$, M7.7$^f$            & 2800$^b$           & MBO80, Sp~72        \\
S-8    &  03 29 03.39 & +31 18 39.9 & 15.833 & 14.000 & M8.2$^a$, M8.5$^b$, M7.4$^c$, M8.4$^f$  & 2850$^a$, 2600$^b$ & MBO88, ASR63, Sp~80 \\  
S-9    &  03 29 05.54 & +31 10 14.2 & 17.072 & 15.667 & M8$^b$, M8.6$^f$	                & 2600$^b$	     &                     \\ 
S-10   &  03 29 05.66 & +31 20 10.7 & 17.113 & 15.485 & 	                                & 2500$^b$           & MBO143, Sp~86             \\ 
S-11   &  03 29 07.17 & +31 23 22.9 & 17.87  & 15.62  & M9.2$^f$                                & (2600)$^b$         & MBO141              \\ 
S-12   &  03 29 09.33 & +31 21 04.2 & 16.416 & 13.150 & 	                                & (2500)$^b$         & MBO70, Sp~93               \\ 
S-13   &  03 29 10.79 & +31 22 30.1 & 14.896 & 12.928 & M7.5$^b$, M7.4$^c$, M8.1$^f$            & 3000$^b$           & MBO62               \\ 
S-14   &  03 29 14.43 & +31 22 36.2 & 14.606 & 13.035 & M7$^b$, M6.6$^c$, M7.7$^f$              & 2900$^b$           & MBO66               \\ 
S-15   &  03 29 17.76 & +31 19 48.1 & 14.803 & 12.988 & M7.5$^b$, M6.5$^c$, M7.8$^f$            & 3000$^b$           & MBO64, ASR80, Sp~112\\ 
S-16   &  03 29 28.15 & +31 16 28.5 & 13.054 & 12.091 & M8.5$^a$, M7.5$^b$, M7.5$^e$, M9.1$^f$  & 2850$^a$, 2600$^b$, 2761$^e$ & Sp~164    \\ 
S-17   &  03 29 33.87 & +31 20 36.2 & 16.562 & 15.481 & M7.5$^a$, M8$^b$, M8.7$^f$              & 2600$^a$, 2500$^b$ & MBO140              \\ 
S-18   &  03 29 35.71 & +31 21 08.5 & 18.50  & 16.94  & M8.2$^f$                                & 2500$^b$           & Sp~129              \\ 
S-19   &  03 29 36.36 & +31 17 49.8 & 17.91  & 16.38  & 	                                & 2700$^b$           &                     \\ 
S-21   &  03 28 47.34 & +31 11 29.8 & 15.484 & 12.702 & 	                                & 3100$^b$           & ASR117              \\ 
\hline
ASR15  &  03 28 56.94 & +31 15 50.3 & 15.056 & 13.461 & M7.4$^c$, M6.0$^d$                      &	             &                     \\
ASR17  &  03 28 57.15 & +31 15 34.5 & 15.405 & 13.186 & M7.4$^c$, M6.0$^d$                      &	             & Sp~68               \\ 
MBO73  &  03 28 58.24 & +31 22 09.3 & 16.004 & 13.367 & M6.4$^c$                                &	             & Sp~70               \\ 
ASR24  &  03 29 11.30 & +31 17 17.5 & 13.977 & 12.915 & M8.2$^c$, M8.0$^d$                      &	             &                     \\ 
MBO69  &  03 29 24.45 & +31 28 14.9 & 14.041 & 12.686 & M7.0$^a$, M7.4$^c$                      & 3200$^a$           &                     \\
\hline
ASR29  &  03 29 13.61 & +31 17 43.4 & 16.441 & 13.028 &  M5$^d$                                 & 3250$^a$           &                     \\	
ASR105 &  03 29 04.66 & +31 16 59.1 & 15.550 & 12.665 &  M6$^d$                                 &	             & Sp~84               \\ 
ASR8   &  03 29 04.06 & +31 17 07.5 & 13.310 & 12.313 &  M7$^d$                                 &	             &                     \\ 
MBO78  &  03 29 00.15 & +31 21 09.2 & 16.466 & 13.349 &  M5$^d$                                 &	             & Sp~75               \\
\hline
Sp~45  &  03 28 43.55 & +31 17 36.4 & 12.219 & 10.138 &  M5.0$^e$                               & 3125$^e$           & ASR127              \\
Sp~46  &  03 28 44.07 & +31 20 52.8 & 14.245 & 12.627 &  M7.3$^a$, M7.5$^e$                     & 3050$^a$, 2829$^e$ &                     \\
Sp~49  &  03 28 47.82 & +31 16 55.2 & 12.940 & 10.909 &  M8.0$^e$                               & 2710$^e$           & ASR111              \\ 
Sp~53  &  03 28 52.13 & +31 15 47.1 & 13.161 & 12.029 &  M7.0$^e$                               & 2846$^e$           & ASR45               \\ 
Sp~55  &  03 28 52.90 & +31 16 26.4 & 13.616 & 12.476 &  M5.0$^e$                               & 3154$^e$           & ASR46               \\ 
Sp~58  &  03 28 54.07 & +31 16 54.3 & 13.027 & 11.599 &  M5.0$^e$                               & 3098$^e$           & ASR42               \\ 
Sp~94  &  03 29 09.48 & +31 27 20.9 & 14.154 & 12.692 &  M5.0$^e$                               & 3098$^e$           & MBO60               \\ 
Sp~105 &  03 29 13.03 & +31 17 38.3 & 15.231 & 14.158 &  M8.0$^e$                               & 2710$^e$           & ASR28               \\ 
Sp~131 &  03 29 37.73 & +31 22 02.4 & 13.987 & 12.958 &  M7.6$^a$, M7.0$^e$                     & 2850$^a$, 2891$^e$ & MBO65               \\ 
Sp~157 &  03 29 12.79 & +31 20 07.7 & 14.676 & 13.294 &  M7.6$^a$, M7.7$^c$, M7.5$^e$           & 2950$^a$, 2812$^e$ & MBO75, ASR83        \\ 
Sp~177 &  03 29 24.83 & +31 24 06.2 & 14.433 & 13.383 &  M6.5$^a$, M6.7$^c$, M6.5$^e$  	        & 3000$^a$, 2957$^e$ & MBO77               \\
Sp~71  &  03 28 58.24 & +31 22 02.1 & 14.912 & 12.406 &  M6$^e$                                 & 2990$^e$           & MBO47               \\
\enddata
\tablenotetext{1}{photometry from 2MASS (most objects) or SONYC (for SONYC-NGC1333-1, 4, 6, 11, 18, 19)}
\tablenotetext{2}{spectral types or effective temperatures from the following references: a - this paper, 
b - \citet{2009ApJ...702..805S}, c - \citet{2004AJ....127.1131W}, d - \citet{2007AJ....133.1321G}, 
e - \citet{2009AJ....137.4777W}, f - HPI spectral types for spectra in \citet{2009ApJ...702..805S}} 
\tablenotetext{3}{identifiers are from \citet[][ASR]{1994A&AS..106..165A},
\citet[][MBO]{2004AJ....127.1131W}, and the Spitzer survey by \citet[][Sp]{2008ApJ...674..336G}}							       
\end{deluxetable*}		

\subsection{The number of brown dwarfs}
\label{s41}

\begin{figure}
\center
\includegraphics[width=6.0cm,angle=-90]{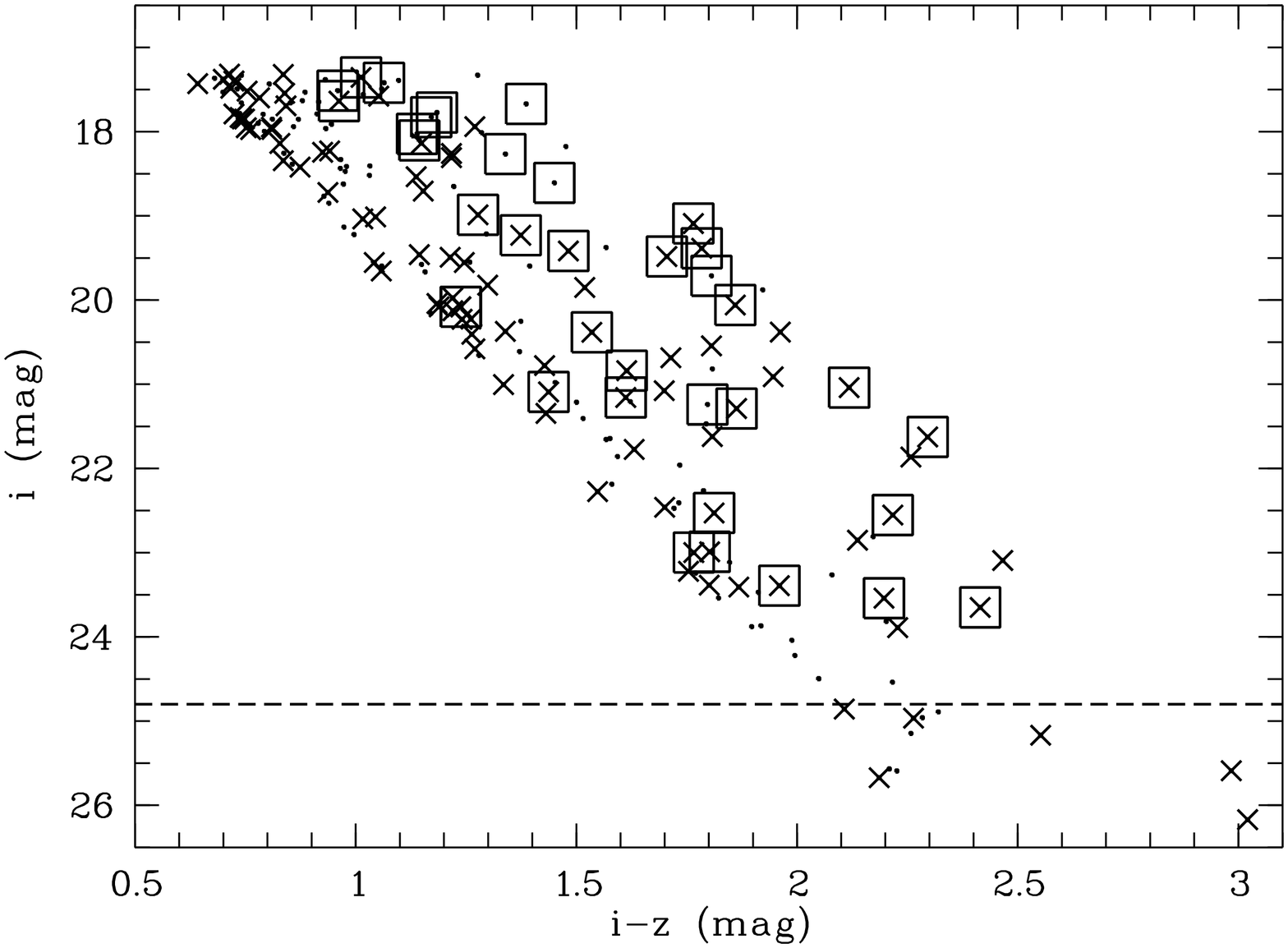} 
\caption{Colour-magnitude diagram for the iz candidates (dots), originally identified in 
\citet{2009ApJ...702..805S}. Crosses are objects for which we have obtained spectra in this paper or 
\citet{2009ApJ...702..805S}. Confirmed very low mass objects from this paper or the literature are 
marked with squares. The horizontal dashed line shows the completeness limit of the survey estimated 
in \citet{2009ApJ...702..805S}. \label{f8}}
\end{figure}

Based on our comprehensive spectroscopy, we can put some constraints on the total number of brown dwarfs
in NGC1333 and the mass limits of the current surveys. For this purpose, we use the iz survey presented in 
\citet{2009ApJ...702..805S}. In Fig. \ref{f8} we plot the iz colour-magnitude diagram for the 196 candidates 
selected in \citet{2009ApJ...702..805S}. The confirmed brown dwarfs (Tables \ref{t1} and \ref{t2}), either by us 
or other groups, are marked with squares; all objects for which we have obtained spectroscopy with crosses. Note 
that the iz candidates are selected only with a cut in colour and a cut in PSF shape to rule out extended objects. 
No other selection criteria have been used, i.e. this sample is as unbiased as possible.

We have useful spectra for 98/196 candidates; out of these 98, 24 are confirmed by our spectra. In total
the iz sample contains 35 confirmed objects with $A_V \lesssim 12$\,mag. Thus, we have a yield of 24/98 (24\%) and 
would expect to find 24 more objects among the candidates for which we do not have spectra. Since 11 of them
(35 minus 24) have already been confirmed by other groups, the expected number of additional very low mass 
objects from this iz selection is 13. 

The low-mass end of the diagram deserves additional discussion. The faintest confirmed brown dwarfs in Fig. 
\ref{f8} are SONYC-NGC1333-1, 30, and 36 at $i=23.4-23.7$. Comparing their effective temperatures with the 
1\,Myr DUSTY and COND isochrones, it seems likely that they have masses of $\lesssim 0.02\,M_{\odot}$. 
If our estimate of $T_{\mathrm{eff}} = 2000$\,K for SONYC-NGC1333-36 is correct (Sect. \ref{s34}), the 
best mass estimate would be in the range of 0.006$\,M_{\odot}$. 

We have taken spectra for 7 fainter objects, but none of them is a brown dwarf, which might indicate that we 
have reached the 'bottom of the IMF', as preliminarily stated in \citet{2009ApJ...702..805S}. However we may not
be 100\% complete in this magnitude range. The formal completeness limits of the iz survey at $i=24.8$\,mag, 
determined from the peak of the histogram of the magnitudes \citep[see][]{2009ApJ...702..805S}, is shown with a 
dashed line. This limit has been derived for a field of view of $34' \times 27'$, but most of the cluster members 
are located in a smaller region of $10' \times 10'$ which is partially affected by significant 
background emission from the cloud (see Fig. \ref{f50}. Thus, the completeness limit in the relevant areas might 
not always reach the value shown in Fig. \ref{f8}. 					

Thus, based on our new data we retract the previous claim by \citet{2009ApJ...702..805S} stating 
a deficit of objects with $M<0.02\,M_{\odot}$ in NGC1333, for two reasons: a) The updated brown dwarf 
census contains a few of objects with masses at or below 0.02$\,M_{\odot}$, including one with an estimated 
mass of 0.006$\,M{\odot}$. b) The current survey may not be complete at the lowest masses, i.e. 
we cannot exclude the presence of a few more objects with $M<0.02\,M_{\odot}$.

The census for $M>0.02\,M_{\odot}$ is more robust. From the 35 confirmed members in the iz diagram we
subtract 10 which are probably slightly above the substellar boundary (see discussion in Sect. \ref{s4}).
We also subtract the 3 which are likely below $0.02\,M_{\odot}$ and add the 13 which we are still missing. 
This gives a total number of $\sim 35$ brown dwarfs down to $0.02\,M_{\odot}$ and with $A_V\lesssim 12$\,mag.

\subsection{The star vs. brown dwarf ratio}
\label{s42}

As a proxy for the shape of the mass function, previous authors have used the ratio of stars to brown 
dwarfs, where these two groups are defined by a range of masses. These ratios are more robust against 
uncertainties in the masses than a complete IMF. \citet{2008ApJ...683L.183A} use a range of 
0.08-1.0$\,M_{\odot}$ for stars and 0.03-0.08$\,M_{\odot}$ for brown dwarfs, hereafter called $R_1$. 
Other authors use 0.08-10$\,M_{\odot}$ for stars and 0.02-0.08$\,M_{\odot}$ for brown dwarfs, hereafter 
called $R_2$ \citep[e.g.][]{2002ApJ...580..317B,2002ApJ...573..366M,2003ApJ...593.1093L}. Since
the number of high-mass stars is small, the two ratios $R_1$ and $R_2$ should be fairly similar.

The comparison with NGC1333 is complicated by the fact that no comprehensive spectroscopic census is 
available for the stars. The best starting point is probably the Spitzer analysis by \citet{2008ApJ...674..336G}. 
They find a total of 137 Class I and II members with disk, from which 94 are in 2MASS. Objects not detected in 2MASS are 
likely embedded sources with high extinction $A_V>10$\,mag and thus not comparable with our brown
dwarf sample. We calculated absolute J-band magnitudes for this sample using the dereddening described
in Sect. \ref{s32} and assuming a distance of 300\,pc, which gives a range of $M_J = 0-9$\,mag. Comparing
with the BCAH 1\,Myr isochrone \citep{1998A&A...337..403B} the sample contains 13 objects with $M>1.0\,M_{\odot}$,
52 objects with $0.08<M<1.0\,M_{\odot}$ and 29 with $0.02<M<0.08\,M_{\odot}$. These are only objects with 
disks; correcting for a disk fracton of 83\% \citep{2008ApJ...674..336G} shifts the numbers to 16, 63, 35. 
The latter number is consistent with the estimate of brown dwarfs in this cluster given in Sect. \ref{s41}. 
Out of 35 brown dwarfs, the number of objects with masses above 0.03$\,M_{\odot}$ would be 28.

Based on these estimates the ratios for NGC1333 become $R_1 = 63/28 = 2.3\pm 0.5$ and 
$R_2 = 79/35 = 2.3 \pm 0.5$ (see below for an explanation of the uncertainties). Our value 
for $R_1$ is somewhat larger than our first estimate given in \citet{2009ApJ...702..805S} of 
$1.5\pm 0.3$, mainly because we use here the cutoff at 0.03$\,M_{\odot}$ to be consistent 
with \citet{2008ApJ...683L.183A}.

The uncertainties for $R_1$ and $R_2$ stated above are 1$\sigma$ confidence intervals and have
been derived based on the prescription provided by \citet{2011PASA...28..128C}. This prescription
is given for population proportions ('success counts'). Therefore, we use the Cameron equation to calculate 
the confidence intervals $\sigma_{k1}$ for the ratio of number of stars to the sum of stars and brown 
dwarfs ($k1$) and $\sigma_{k2}$ for the ratio of number of brown dwarfs to the same sum ($k2$). 
The confidence intervals for $R_1$ and $R_2$ are then derived as follows:
\begin{equation}
R_{\mathrm{max}} = \frac{k1 + \sigma_{k1}}{k2 - \sigma_{k2}}
\end{equation}
\begin{equation}
R_{\mathrm{min}} = \frac{k1 - \sigma_{k1}}{k2 + \sigma_{k2}}
\end{equation}

\begin{deluxetable}{lll}
\tabletypesize{\scriptsize}
\tablecaption{Star to brown dwarf ratios for various star forming regions (see text)
\label{t4}}
\tablewidth{0pt}
\tablehead{
\colhead{Region} & \colhead{$R_1$\tablenotemark{a}} & \colhead{$R_2$\tablenotemark{a}}}
\tablecolumns{3}
\startdata
NGC1333          & 2.3\tablenotemark{b} (1.8-2.8)  & 2.3\tablenotemark{b} (1.8-2.7) \\
ONC              & 3.3\tablenotemark{c} (2.8-3.9)  & 3.8,4.5,5.0\tablenotemark{f,g,h}   \\ 
UpSco            & 3.8\tablenotemark{d} (3.1-4.5)  & --                \\
NGC2024          & 3.8\tablenotemark{c} (2.6-5.2)  & --                \\
Chamaeleon       & 4.0\tablenotemark{c} (2.3-6.0)  & 3.9\tablenotemark{i} (2.9-5.0)\\ 
$\rho$-Oph       & 5.1\tablenotemark{e} (3.8-6.4)  & 4.8\tablenotemark{e} (3.7-6.0) \\
Taurus           & 6.0\tablenotemark{c} (4.5-7.7)  & 7.3\tablenotemark{f} (5.1-9.6)\\ 
IC348            & 8.3\tablenotemark{c} (6.4-10.5) & 8.0\tablenotemark{f} (6.3-10.0)
\enddata
\tablenotetext{a}{In brackets 1$\sigma$ confidence intervals, see text. Note that for the $R_2$ value
in the ONC no absolute numbers are available, thus no error estimate is possible. For $\rho$-Oph we
adopted the average numbers from the ranges given by Muzic et al. (subm.).}
\tablenotetext{b}{this paper}
\tablenotetext{c}{\citet{2008ApJ...683L.183A}}
\tablenotetext{d}{\citet{2011arXiv1108.1309D}} 
\tablenotetext{e}{Muzic et al. (subm.)}
\tablenotetext{f}{\citet{2003ApJ...593.1093L}}
\tablenotetext{g}{\citet{2002ApJ...573..366M}}
\tablenotetext{h}{\citet{2004ApJ...610.1045S}}
\tablenotetext{i}{\citet{2007ApJS..173..104L}}
\end{deluxetable}

\begin{figure}
\center
\includegraphics[width=6.0cm,angle=-90]{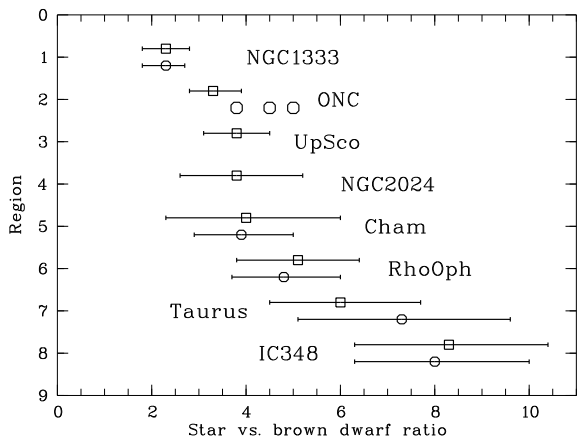} 
\caption{Star vs. brown dwarf ratios for various star forming regions, from Table \ref{t4}. $R_1$ is plotted
with squares, $R_2$ with circles. The number on the y-axis is identical with the row in Table \ref{t4}.
\label{f40}}
\end{figure}

In Table \ref{t4} we compare the ratios for NGC1333 with the available literature values for $R_1$ 
and $R_2$ for other regions. The same numbers are plotted in Fig. \ref{f40} for illustration. To have 
accurate and consistent confidence intervals, we re-calculated the errors for all literature values 
as described above. NGC1333 has the lowest ratios measured so far in any star forming 
region, suggesting that the number of brown dwarfs in NGC1333 is unusually high, which is in line with 
the conclusion in \citet{2009ApJ...702..805S}. In particular, the ratios for NGC1333 deviate by more
than 2$\sigma$ from those in IC348. It should be noted that the current census for IC348 
\citep{2003ApJ...593.1093L} is nearly complete down to 0.03$\,M_{\odot}$ and covers most of the cluster, 
which makes the difference to NGC1333 even more striking.

This finding has to be substantiated with further survey work in diverse regions. In
Table \ref{t4} we list the statistical 1$\sigma$ confidence intervals, purely based on the sample
sizes. These statistical errors do not take into account additional sources of  
uncertainty, e.g. unrecognised biases, inconsistencies in sample selection or problems
with the mass estimates, i.e. the actual errors may be larger than listed in Table \ref{t4}.
In particular, it is important to note that all mass estimates are necessarily model-dependent. 
For the value in NGC1333 we use the BCAH isochrones, mainly because they cover the brown dwarf
regime down to the Deuterium burning limit. The problems and uncertainties of these type
of models at very young ages are well-documented \citep{2002A&A...382..563B}.

The best way of assessing the overall uncertainties is to compare results from independent 
surveys. As can be seen in Table \ref{t4}, so far the results from independent groups agree
within the statistical errorbars, with the possible exception of the ONC.\footnote{A new paper
by \citet{2011arXiv1108.5581A} updates the value of for the ONC to $R_1 = 2.4 \pm 0.4$, based
on an HST survey covering a larger area than previous studies.} Such an independent
confirmation is required for NGC1333 as well. 

If confirmed, the unusually low ratio of stars to brown dwarfs in NGC1333 could point to regional 
differences in this quantity, possibly indicating environmental differences
in the formation of very low mass objects. One option to explain this is turbulence, as very low 
mass cores which can potentially collapse to brown dwarfs could be assembled by the turbulent flow
in a molecular cloud \citep{2004ApJ...617..559P}. At first glance this could be a realistic possibility 
for NGC1333, where the cloud is strongly affected by numerous outflows \citep{2005ApJ...632..941Q}, although
it is not clear if the turbulence in NGC1333 is mainly driven by these outflows \citep{2009ApJ...707L.153P}. 
Alternatively, additional brown dwarfs could form by gravitational fragmentation of gas falling into the 
cluster center \citep{2008MNRAS.389.1556B}. This latter mechanism would benefit from the fact that NGC1333 
has a higher stellar density and thus a stronger cluster potential than most other nearby star forming 
regions \citep{2009ApJ...702..805S}.

\subsection{Spatial distribution}
\label{s44}

\begin{figure}
\center
\includegraphics[width=6.0cm,angle=-90]{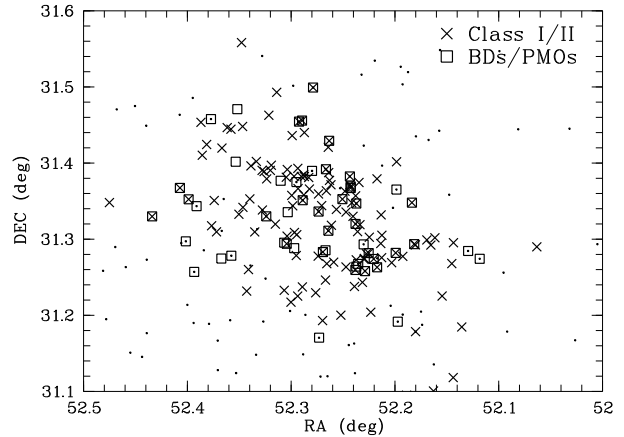} 
\caption{Spatial distribution of NGC1333 members. Crosses are all 137 objects with Spitzer excess by 
\citet{2008ApJ...674..336G}, squares are all confirmed brown dwarfs (Tables \ref{t1} and \ref{t2}). 
Objects with spectroscopy for which we can exclude that they are substellar members are shown with dots.
\label{f9}}
\end{figure}

\begin{figure}
\center
\includegraphics[width=6.0cm,angle=-90]{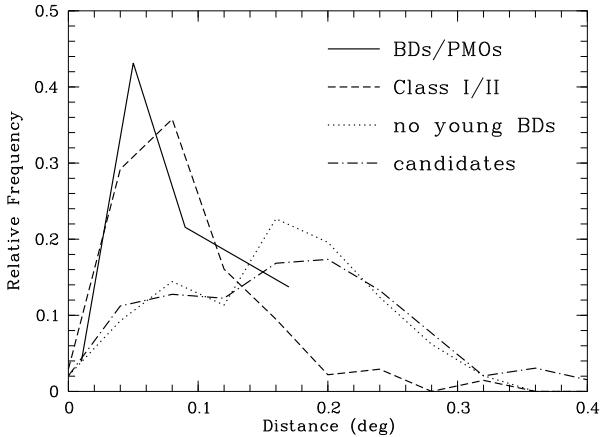} 
\caption{Histogram of the distance from the cluster center for confirmed brown dwarfs (solid line,
Tables \ref{t1} and \ref{t2}), objects with Spitzer excess by \citet{2008ApJ...674..336G} (dashed line), 
objects for which we can exclude that they are substellar members (dotted line), and all IZ 
candidates (dash-dotted line). As cluster center we used the average position of the Class I/II sources.
\label{f30}}
\end{figure}

In Fig. \ref{f9} we show the spatial distribution of the sample of very low mass objects listed in
Tables \ref{t1} and \ref{t2} (squares). For comparison, the positions of the 137 Class I and Class II sources 
\citep{2008ApJ...674..336G} are overplotted with crosses. The dots indicate the positions of all targets 
for which we have obtained spectra but which are not confirmed as very low mass objects. Additionally,
we show the frequency of objects as a function of distance from the cluster center in Fig. \ref{f30}, 
again for the same three samples, and in addition for all photometric candidates from 
our iz catalogue (dash-dotted line).

In the two figures, the spatial distribution of brown dwarfs is strongly clustered and indistinguishable 
from the distribution of the total population of Class I/II sources in NGC1333. For the two samples, 
the average $\alpha$ and $\delta$ differ only by 0.4' and 0.3', respectively, which is $<10$\% of the cluster 
radius. Adopting the average position of the Class I/II sources as cluster center, the average distance from 
the center is similar in the two samples, 5.2' for the very low mass objects and 5.5' for the Class I/II 
sources. The fraction of objects with distance from the cluster center of $d<0.1$, $0.1<d<0.2$, $d>0.2$\,deg 
is 65, 35, 0\% for the very low mass objects and 67, 26, 6\% for the Class I/II objects. The scatter in the 
positions is $\sigma_{\alpha} = 0.071$ and $\sigma_{\delta} = 0.069$\,deg for the very low mass objects, 
and $\sigma_{\alpha} = 0.067$ and $\sigma_{\delta} = 0.085$\,deg for the Class I/II sources. For all these 
quantities there are no significant differences between the two samples.

The figures also show that our spectroscopic follow-up covers an area that is about 1.5-2 as large (in radius) 
than the cluster itself. We took spectra for 31 candidates with distances of $>0.2$\,deg from the 
cluster center, but none of them turned out to be a brown dwarf. There are still 43 candidates from the IZ
photometry outside 0.2\,deg (see Sect. \ref{s41}) for which we do not have spectra, but based on our current 
results, it is unlikely that they contain any very low mass cluster members. Thus, our wide-field follow-up 
spectroscopy shows that there is no significant population of brown dwarfs at $d>0.2$\,deg from the cluster 
center, corresponding to $\sim 1$\,pc at a distance of 300\,pc. 

It has been suggested that gravitational ejection occurs at an early stage in the evolution of substellar 
objects, either from multiple stellar/substellar systems \citep{2001AJ....122..432R,2005ApJ...623..940U} or from 
a protoplanetary disk \citep{2009MNRAS.392..413S}. This ejection is thought to remove the objects from their 
accretion reservoir and thus sets their masses. In these scenarios one could expect the brown dwarfs to have 
high spatial velocities in random directions.

An ejection velocity of 1\,kms$^{-1}$ would allow the object to travel 1\,pc in 1\,Myr, i.e. in 
the case of NGC1333 this would be sufficient to reach the edge of the cluster. However, the gravitational 
potential of the cluster will significantly brake the motion of the brown dwarf. Assuming a total cluster 
mass of 500$\,M_{\odot}$ \citep{1996AJ....111.1964L} homogenuously distributed in a sphere with 1\,pc 
radius, a brown dwarf that gets ejected in the cluster center with 1.5, 2, 3\,kms$^{-1}$ would reach a 
velocity of approximately 0.5, 1.4, 2.6\,kms$^{-1}$ at a distance of 1\,pc from the center. All objects with 
ejection velocities of $\gtrsim 2$\,kms$^{-1}$ would have moved to distances significantly larger than 
1\,pc over 1\,Myr. As shown above, the presence of such objects can be excluded from our data.

The scenarios by \citet{2005ApJ...623..940U} and \citet{2009MNRAS.392..413S} predict that a substantial
fraction of ejected brown dwarfs (more than 50\% in some simulations) exceed this velocity threshold of
$2$\,kms$^{-1}$. These models would require some tuning to reproduce a spatial distribution as observed in 
NGC1333. However, such simplified scenarios do not take into account that dynamical interactions affect 
the total cluster population, not exclusively the brown dwarfs. The cluster formation simulations by 
\citet{2009MNRAS.392..590B} show that the velocity dispersion in a dense cluster is not expected
to increase in the very low mass regime. Although brown dwarfs undergo ejection in the simulations, this 
does not lead to a velocity offset in comparison to the stars. NGC1333 seems to be consistent with this 
picture.

As a side comment, we note that the parameters in the main simulation in \citet{2009MNRAS.392..590B}
with gas mass of 500$\,M_{\odot}$ and cloud radius of 0.4\,pc are fairly similar to the properties of 
NGC1333, although the simulation produces a much higher number of stars and brown dwarfs (total stellar 
mass of 191$\,M_{\odot}$ vs. $\sim 50\,M_{\odot}$  in NGC1333). 

\subsection{Disks}
\label{s45}

\begin{figure}
\center
\includegraphics[width=6.0cm,angle=-90]{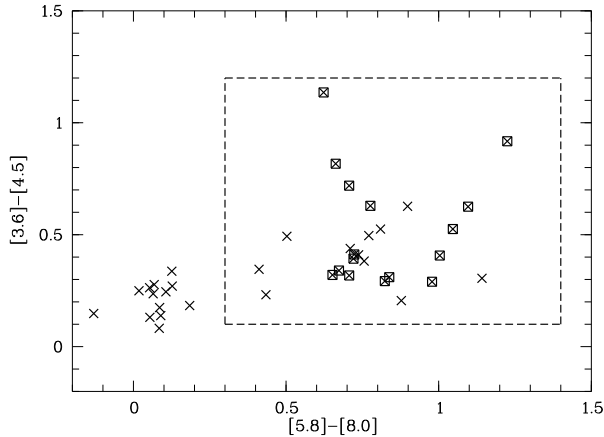} 
\caption{Spitzer/IRAC colour-colour diagram for the very low mass sources listed in Tables \ref{t1} and 
\ref{t2}, using the 'HREL' photometry from the C2D survey. We only plot objects with photometric errors $<40$\% in all
four bands (41 out of 51). The position of the Class II objects from the survey by \citet{2008ApJ...674..336G}
is indicated by the dashed box. Objects with reliable detection at 24$\,\mu m$ are marked with squares
\label{f10}}
\end{figure}

In Fig. \ref{f10} we plot the Spitzer/IRAC colour-colour diagram for the sample listed in  
Tables \ref{t1} and \ref{t2}, again based on the C2D-'HREL' catalogue. Out of the sample of 51 sources 
with confirmed spectral type M5 or later, 41 have photometric errors $<40$\% in all four IRAC bands and 
are shown in this plot. The figure shows the typical appearance with two groups, one around the 
origin, the second with significant colour excess in mid-infrared bands due to the presence of 
circum-(sub)-stellar material. In this sample of 41 objects, 27 show evidence for a disk, i.e. 
$66\pm 8$\%. All of them have colours comparable to the Class II sources identified in 
\citet{2008ApJ...674..336G}.

The derived disk fraction of 66\% is only valid for the sample of 41 objects with reliable Spitzer
detection. In the entire sample of 51 very low mass sources in NGC1333 listed in Tables \ref{t1} and \ref{t2}, 
the disk fraction is likely to be smaller, because the ten objects which are not detected by Spitzer
are unlikely to have a disk. Correcting for this effect, the disk fraction in the full sample
could be as low as 27/51 or 55\%. Therefore, we consider the disk fraction of 66\% to be an upper 
limit.

For comparison, for the total cluster population \citet{2008ApJ...674..336G} derive a disk fraction 
of 83\% from a Spitzer survey. This number has been derived for objects with $K<14$\,mag. This magnitude 
limit was chosen by \citet{2008ApJ...674..336G} because it corresponds to $M=0.08\,M_{\odot}$ at age of 
1\,Myr and $A_V=20$\,mag. Their sample thus includes mostly stars, but also some brown dwarfs (as evident 
from Table \ref{t2}, which contains a number of objects from the \citet{2008ApJ...674..336G} sample, 
marked with 'Sp'.) The Spitzer sample contains a substantial number of objects with $A_V>10$\,mag, which 
are rare among the currently known brown dwarfs. It is possible that some of the heavily embedded 
brown dwarfs with $A_V>10$\,mag have not been found yet. This could explain the discrepancy in the
disk fractions.

Our disk fraction is consistent with the values derived for very low mass members in $\sigma$\,Ori, 
Chamaeleon-I, and IC348 \citep{2008ApJ...688..362L} although all three regions are thought to be 
somewhat older (2-3\,Myr) than NGC1333. 

A more detailed SED analysis was carried out for objects with an additional datapoint at 24$\,\mu m$.
19 of the objects in Fig. \ref{f10} have MIPS fluxes at 24$\,\mu m$ with errors $<40$\%. At this wavelength
the images are strongly affected by the cloud emission and blending. To make sure that the fluxes are
trustworthy, we checked all objects in a 24$\,\mu m$ image obtained in the 
Spitzer program \#40563 (PI: K. Wood, AOR 23712512), which is deeper than the C2D mosaics. 
After visual inspection, 3 objects were discarded; the remaining 16 are point sources at
24$\,\mu m$ and are marked with squares in Fig. \ref{f10}. 

\begin{figure}
\center
\includegraphics[width=6.0cm,angle=-90]{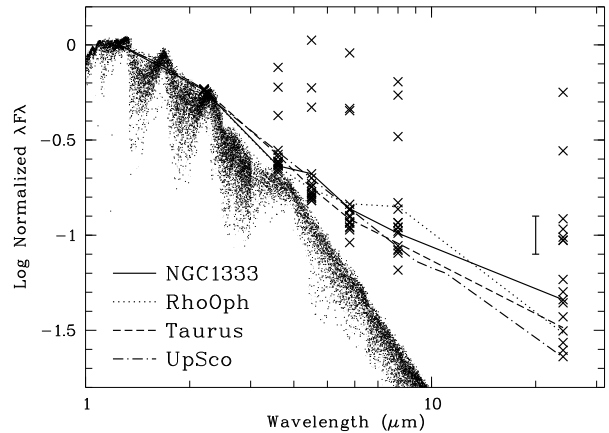} 
\caption{Spectral energy distributions for 16 very low mass objects with disks in NGC1333 (crosses). The 
SEDs have been dereddened and scaled to the J-band fluxes. We overplot the typical SED for NGC1333 
(solid line), $\rho$\,Oph (dotted line) Taurus (dashed line), and UpSco (dash-dotted line). 
We also show the photospheric SED from the 2800\,K DUSTY-AMES model \citep{2001ApJ...556..357A} with 
small dots. A typical errorbar for the 24$\,\mu$ fluxes is overplotted.
\label{f11}}
\end{figure}

In Fig. \ref{f11} we show their SEDs after dereddening (see Sect. \ref{s32}) and scaling to the J-band
flux (crosses). For comparison, the photospheric SED from a model spectrum is overplotted with small dots.
To assess the disk evolution in the substellar regime, we derive the typical SED for 
NGC1333 and three other star forming regions: $\rho$\,Oph (1\,Myr), Taurus (2\,Myr), and UpSco (5\,Myr). 
For this purpose we selected the objects which are detected in all four IRAC bands and at 24$\,\mu m$. 
For $\rho$\,Oph we started with the census in Muzic et al. (subm.) and made use of the C2D data. 
For Taurus we used published Spitzer data from \citet{2007A&A...465..855G} and \citet{2006ApJ...645.1498S}. 
For UpSco the data from \citet{2007ApJ...660.1517S} was used. When comparing the SEDs from different regions, 
one has to take into account that the depth of the 24$\,\mu m$ observations is not the same; thus the median 
SED is affected by incompleteness at low flux levels. Instead we plot the SED for the object that has the 
10th highest flux level at 24$\,\mu m$ after converting to $\lambda F_\lambda$ and scaling to the J-band flux. This
represents an estimate for a typical SED unaffected by the depth of the Spitzer observations and
the distance to the cluster. Note that all objects used for this exercise are spectroscopically 
confirmed members of the respective clusters.

For wavelengths $<8\,\mu m$ the four median SEDs are fairly similar. At 8$\,\mu m$ the SEDs in the youngest 
regions (NGC1333, $\rho$\,Oph) are slightly enhanced. The biggest differences are seen at 24$\,\mu m$,
particularly when comparing NGC1333 with UpSco. This is mostly due to the fact that NGC1333 harbours a
few objects with unusually strong excess emission, which are missing in UpSco 
\citep[compare with Fig. 1 in][]{2007ApJ...660.1517S}, indicating that the objects in NGC1333 are in an
early evolutionary stage compared with the other regions. As demonstrated in \citet{2009MNRAS.398..873S} 
a large spread in 24$\,\mu m$ fluxes, as seen in NGC1333, can easily be explained by a range of flaring 
angles in the disks.

\section{Conclusions}
\label{s5}

As part of our survey program SONYC, we present a census of very low mass objects in the young cluster
NGC1333 based on new follow-up spectroscopy from Subaru/FMOS. To derive reliable spectral types
from our data, we define a new spectral index based on the slope of the H-band peak. We find 10 new 
likely brown dwarfs in this cluster, including one with a spectral type $\sim$L3 and two more with 
spectral type around or later than M9. These objects have estimated masses of 0.006 to 0.02$\,M_{\odot}$,
the least massive objects identified thus far in this region. This demonstrates that the mass function
in this cluster extends down to the Deuterium burning limit and beyond. By combining the findings from
our SONYC survey with results published by other groups, we compile a sample of 51 objects with spectral
types of M5 or later in this cluster, more than half of them found by SONYC. About 30-40 of them are likely 
to be substellar. The star vs. brown dwarf ratio in NGC1333 is significantly lower than in other nearby 
star forming regions, possibly indicating environmental differences in the formation of brown dwarfs. 
We show that the spatial distribution of brown dwarfs in NGC1333 closely follows the distribution of 
the stars in the cluster. The disk fraction in the brown dwarf sample is $<66$\%, lower than for the 
stellar members, but comparable to the brown dwarf disk fraction in 2-3\,Myr old regions. The substellar 
members in NGC1333 show a large fraction of highly flared disks, evidence for the early evolutionary 
state of the cluster.

\acknowledgments
The authors would like to thank the Subaru staff, especially Dr. Naoyuki Tamura and Dr. 
Kentaro Aoki, for the assistance during the observations and their preparation. We are 
grateful to Ms. Yuuki Moritani, Mr. Kiyoto Yabe and Prof. Fumihide Iwamuro (Kyoto 
University) for their help with the FMOS data reduction. We thank the anonymous
referee for a constructive report that helped to improve the paper.
This work made use of results from the Spitzer program \#40563; we thank Jane Greaves, 
Chris Poulton, and Kenneth Wood from the University of St. Andrews for their help in the 
framework of this campaign. We also thank Ewan Cameron for discussing the calculation of
confidence intervals and David Lafreni\`ere for making his spectra available to us. AS 
acknowledges financial support the grant 10/RFP/AST2780 from the Science Foundation 
Ireland. The research was supported in large part by grants from the Natural Sciences 
and Engineering Research Council (NSERC) of Canada to RJ. This research has benefitted 
from the SpeX Prism Spectral Libraries, maintained by Adam Burgasser at 
{\tt http://pono.ucsd.edu/\textasciitilde adam/browndwarfs/spexprism/}.

\appendix

\section{Spectroscopically excluded objects}
\label{a1}

In Tables \ref{t10} and \ref{t11} we provide a full list of objects for which we obtained spectra 
and which were not classified as young very low mass objects based on the shape of their near-infrared 
spectrum (see Sect. \ref{s31}). The spectra come from the first campaign with 
MOIRCS \citep{2009ApJ...702..805S} and from the second run with FMOS (this paper). 
Most of these objects are likely to be either young stellar objects in NGC1333 
or background stars with effective temperatures above $\sim 3500$\,K or spectral type earlier 
than $\sim$M3. In Table \ref{t10} we also give the J- and K-band photometry
from 2MASS and the identifiers from the photometric surveys by \citet{1996AJ....111.1964L} and
\citet{2004AJ....127.1131W}, if available. Objects without listed identifiers are not known to
have a counterpart within 1". 

\begin{deluxetable*}{llccccl}
\tabletypesize{\scriptsize}
\tablecaption{Objects excluded by spectroscopy in this paper \label{t10}}
\tablewidth{0pt}
\tablehead{
\colhead{$\alpha$(J2000)} & \colhead{$\delta$(J2000)} & \colhead{J (mag)\tablenotemark{1}} & 
\colhead{K (mag)\tablenotemark{1}} & \colhead{Sel\tablenotemark{2}} & \colhead{Spec\tablenotemark{3}}
& \colhead{Identifier\tablenotemark{4}} }
\tablecolumns{7}
\startdata
 03 29 18.71 & +31 32 26.4 & 16.722 & 13.839 & IZ & F &  \\  
 03 29 52.35 & +31 28 13.7 & 14.919 & 13.829 & IZ & F &  \\  
 03 29 48.12 & +31 28 29.4 & 15.550 & 13.663 & IZ & F &  \\  
 03 29 37.80 & +31 27 48.4 & 17.274 & 15.249 & IZ & F &  \\  
 03 29 34.76 & +31 29 08.1 & 13.647 & 11.532 & IZ & F & MBO 43 \\  
 03 29 11.53 & +31 30 05.6 & 16.740 & 13.786 & IZ & F & [LAL96] 241, 242 \\  
 03 28 55.74 & +31 30 58.0 & 15.175 & 13.390 & IZ & F & [LAL96] 143 \\  
 03 28 52.66 & +31 32 04.3 & 16.563 & 14.730 & IZ & F &  \\  
 03 28 46.67 & +31 31 35.4 & 15.647 & 13.933 & IZ & F &  \\  
 03 28 37.55 & +31 32 54.5 & 15.107 & 14.232 & IZ & F &  \\  
 03 28 23.84 & +31 32 49.3 & 14.714 & 13.655 & IZ & F &  \\  
 03 28 44.96 & +31 31 09.9 & 17.057 & 15.211 & IZ & F &  \\  
 03 28 46.24 & +31 30 12.1 & 15.089 & 14.069 & IZ & F & [LAL96] 88 \\  
 03 29 45.53 & +31 26 56.5 & 14.870 & 14.023 & IZ & F &  \\  
 03 30 05.41 & +31 25 13.1 & 14.921 & 13.680 & IZ & F &  \\  
 03 28 49.48 & +31 25 06.6 & 15.957 & 13.658 & IZ & F & MBO 82 \\  
 03 29 10.46 & +31 23 34.8 & 15.633 & 12.762 & JK & F & [LAL96] 231 \\  
 03 28 43.17 & +31 26 06.1 & 15.617 & 13.384 & IZ & F & [LAL96] 78 \\  
 03 28 37.75 & +31 26 32.8 & 16.847 & 14.340 & IZ & F & [LAL96] 60 \\  
 03 27 56.27 & +31 27 00.8 & 16.836 & 13.967 & IZ & F &  \\  
 03 28 07.64 & +31 26 42.4 & 15.974 & 13.719 & IZ & F &  \\  
 03 28 19.51 & +31 26 39.5 & 15.100 & 13.687 & IZ & F & [LAL96] 13 \\  
 03 28 40.22 & +31 25 49.1 & 15.637 & 12.911 & IZ & F &  \\  
 03 28 47.64 & +31 24 06.2 & 14.199 & 11.660 & JK & F & [LAL96] 93 \\  
 03 28 55.22 & +31 25 22.4 & 14.735 & 12.550 & IZ & F & [LAL96] 139 \\  
 03 29 03.32 & +31 23 14.8 & 17.254 & 14.071 & JK & F & [LAL96] 191 \\  
 03 29 03.13 & +31 22 38.2 & 13.724 & 11.323 & JK & F & [LAL96] 189 \\  
 03 29 27.61 & +31 21 10.1 & 14.836 & 13.074 & IZ & F & [LAL96] 324 \\  
 03 29 55.50 & +31 15 30.5 & 15.120 & 14.131 & IZ & F &   \\  
 03 29 52.65 & +31 17 22.9 & 16.325 & 14.970 & IZ & F &  \\  
 03 29 39.61 & +31 17 43.4 &        & 	     & JK & F &  \\  
 03 28 35.46 & +31 21 29.9 & 16.052 & 14.197 & IZ & F & [LAL96] 49 \\  
 03 28 48.45 & +31 20 28.4 & 16.842 & 14.283 & IZ & F & [LAL96] 100 \\  
 03 29 10.82 & +31 16 42.7 & 15.652 & 13.039 & JK & F & [LAL96] 235 \\  
 03 29 21.42 & +31 15 55.3 & 15.396 & 14.004 & IZ & F & [LAL96] 305 \\  
 03 29 45.41 & +31 16 23.2 & 15.118 & 13.767 & IZ & F & \\  
 03 29 50.23 & +31 15 47.9 & 16.980 & 15.106 & IZ & F & \\  
 03 30 01.93 & +31 10 50.5 & 14.374 & 12.921 & IZ & F & \\  
 03 29 54.78 & +31 11 41.7 & 16.679 & 15.253 & IZ & F & \\  
 03 29 34.57 & +31 11 23.8 & 16.750 & 14.643 & IZ & F & [LAL96] 351 \\  
 03 29 36.00 & +31 12 49.6 & 15.522 & 14.615 & IZ & F & [LAL96] 353 \\  
 03 29 26.11 & +31 11 36.9 & 14.781 & 12.874 & IZ & F & [LAL96] 320 \\  
 03 28 01.19 & +31 17 36.5 & 15.587 & 14.136 & IZ & F &  \\  
 03 27 52.51 & +31 19 38.8 & 14.973 & 13.662 & IZ & F &  \\  
 03 28 22.90 & +31 15 21.7 & 15.032 & 13.587 & IZ & F & [LAL96] 20 \\  
 03 29 08.71 & +31 12 01.9 &        & 	     & IZ & F &  \\  
 03 29 12.24 & +31 12 20.5 & 16.588 & 14.722 & IZ & F & [LAL96] 254 \\  
 03 29 18.45 & +31 11 30.5 & 16.413 & 15.530 & IZ & F &  \\  
 03 29 31.00 & +31 11 20.1 & 17.052 & 15.265 & IZ & F &  \\  
 03 29 28.99 & +31 10 00.3 & 13.365 & 10.889 & IZ & F & [LAL96] 329 \\  
 03 29 45.40 & +31 10 35.6 & 14.828 & 13.516 & IZ & F &  \\  
 03 29 46.48 & +31 08 43.6 & 15.205 & 14.096 & IZ & F &  \\  
 03 29 49.16 & +31 09 03.7 & 16.143 & 14.537 & IZ & F &  \\  
 03 29 28.95 & +31 07 40.9 & 15.975 & 15.199 & IZ & F & [LAL96] 330 \\  
 03 29 24.73 & +31 07 26.8 & 	    & 	     & IZ & F &  \\  
 03 29 20.11 & +31 08 53.7 & 15.923 & 14.813 & IZ & F &  \\  
 03 29 09.23 & +31 08 55.4 & 16.330 & 14.912 & IZ & F &  \\  
 03 28 57.45 & +31 09 46.5 & 16.623 & 12.738 & IZ & F & [LAL96] 160 \\  
 03 28 06.32 & +31 10 02.0 & 15.731 & 14.579 & IZ & F &  \\  
 03 28 22.07 & +31 10 42.9 & 13.867 & 11.859 & IZ & F & [LAL96] 18 \\  
 03 28 39.01 & +31 08 07.6 & 15.143 & 13.389 & IZ & F & [LAL96] 65 \\  
 03 29 03.60 & +31 07 11.9 & 14.995 & 13.351 & IZ & F & [LAL96] 197  \\  
 03 29 05.23 & +31 07 10.6 & 16.776 & 14.766 & IZ & F &    
\enddata
\tablenotetext{1}{Photometry from 2MASS,if available}
\tablenotetext{2}{Selected from IZ catalogue (IZ), JK plus Spitzer catalogue (JK)}
\tablenotetext{3}{Source of spectrum: M - MOIRCS, F - FMOS}
\tablenotetext{4}{[LAL96] -- \citet{1996AJ....111.1964L}; MBO -- \citet{2004AJ....127.1131W}; if
no identifier is listed, the object does not have a known counterpart within 1"}
\end{deluxetable*}

\begin{deluxetable*}{llccccl}
\tabletypesize{\scriptsize}
\tablecaption{Objects excluded by spectroscopy by \citet{2009ApJ...702..805S} \label{t11}}
\tablewidth{0pt}
\tablehead{
\colhead{$\alpha$(J2000)} & \colhead{$\delta$(J2000)} & \colhead{J (mag)\tablenotemark{1}} & 
\colhead{K (mag)\tablenotemark{1}} & \colhead{Sel\tablenotemark{2}} & \colhead{Spec\tablenotemark{3}}
& \colhead{Identifier\tablenotemark{4}} }
\tablecolumns{7}
\startdata
 03 28 41.72 & +31 11 15.1 & 	    & 	     & IZ & M & \\  
 03 28 41.97 & +31 12 17.2 & 15.863 & 13.822 & IZ & M & \\  
 03 28 46.21 & +31 12 03.4 & 16.933 & 13.526 & IZ & M & [LAL96] 90\\  
 03 28 48.99 & +31 12 45.1 & 17.836 & 14.048 & IZ & M & [LAL96] 103\\  
 03 28 52.10 & +31 16 29.3 & 16.071 & 13.738 & IZ & M & [LAL96] 123\\  
 03 28 57.25 & +31 07 26.0 & 	    & 	     & IZ & M & [LAL96] 159\\  
 03 28 58.68 & +31 09 39.2 & 17.052 & 12.945 & IZ & M & [LAL96] 169\\  
 03 29 00.70 & +31 22 00.8 & 16.236 & 11.764 & IZ & M & [LAL96] 180\\  
 03 29 17.93 & +31 14 53.5 & 16.625 & 14.039 & IZ & M & [LAL96] 287\\  
 03 29 18.66 & +31 20 17.8 & 17.510 & 14.608 & IZ & M & MBO 109\\  
 03 29 19.86 & +31 18 47.7 & 17.205 & 13.321 & IZ & M & [LAL96] 297\\  
 03 29 28.06 & +31 18 39.0 & 15.307 & 12.846 & IZ & M & [LAL96] 327\\  
 03 29 32.20 & +31 17 07.3 & 15.503 & 13.784 & IZ & M & [LAL96] 341\\  
 03 29 37.41 & +31 17 41.6 & 14.886 & 12.761 & IZ & M & [LAL96] 359\\  
 03 29 08.17 & +31 11 54.6 & 17.247 & 14.980 & IZ & M & [LAL96] 217  
\enddata
\tablenotetext{1}{Photometry from 2MASS,if available}
\tablenotetext{2}{Selected from IZ catalogue (IZ), JK plus Spitzer catalogue (JK)}
\tablenotetext{3}{Source of spectrum: M - MOIRCS, F - FMOS}
\tablenotetext{4}{[LAL96] -- \citet{1996AJ....111.1964L}; MBO -- \citet{2004AJ....127.1131W}; if
no identifier is listed, the object does not have a known counterpart within 1"}
\end{deluxetable*}

\bibliography{aleksbib}

\clearpage

\end{document}